\documentclass[%
 aip,jap,
 amsmath,amssymb,
 reprint,%
]{revtex4-1}

\usepackage{xcolor}
\usepackage{graphicx}
\usepackage{dcolumn}
\usepackage{bm}
\usepackage[utf8]{inputenc}
\usepackage[T1]{fontenc}
\usepackage{mathptmx}

\begin{document}
\preprint{AIP/123-QED}

\title{Tailoring the anomalous Hall effect of SrRuO$_3$ thin films by strain: a first principles study}

\begin{abstract}
Motivated by the recently observed unconventional Hall effect
in ultra-thin films of ferromagnetic SrRuO$_3$ (SRO) we investigate the effect of strain-induced oxygen octahedral distortion in the electronic structure and anomalous Hall response of the SRO ultra-thin films by virtue of density functional theory calculations. Our findings reveal that the ferromagnetic SRO films grown on SrTiO$_3$ (in-plane strain of $-$0.47$\%$) have an orthorhombic (both tilting and rotation) distorted structure, and with an increasing amount of substrate induced compressive strain the octahedral tilting angle is found to be suppressed gradually, with SRO films grown on NdGaO$_3$ (in-plane strain of $-$1.7$\%$)  stabilized in the tetragonal distorted structure (with zero tilting).
Our Berry curvature calculations  predict a positive value of the anomalous Hall conductivity  of $+$76\,S/cm at $-$1.7$\%$ strain, whereas it is found to be negative ($-$156\,S/cm) at $-$0.47$\%$ strain. We attribute the found behavior of the anomalous Hall effect to the nodal point dynamics in the electronic structure arising in response to tailoring the oxygen octahedral distortion driven by the substrate induced strain. 
We also calculate  strain-mediated anomalous Hall conductivity as a function of reduced magnetization obtained by scaling down the magnitude of  the exchange field inside Ru atoms  finding good qualitative agreement with experimental observations, which indicates a strong impact of longitudinal thermal fluctuations of Ru spin moments on the anomalous Hall effect in this system.

\end{abstract}       

\author{Kartik Samanta}
\email{k.samanta@fz-juelich.de}
\affiliation{Peter Grunberg Institut and Institute for Advanced Simulation, 
Forschungszentrum J\"{u}lich and JARA, 52425 J\"{u}lich, Germany}
\author{Marjana Ležaić}
\affiliation{Peter Grunberg Institut and Institute for Advanced Simulation, 
Forschungszentrum J\"{u}lich and JARA, 52425 J\"{u}lich, Germany}
\author{Stefan Blügel}
\affiliation{Peter Grunberg Institut and Institute for Advanced Simulation, 
Forschungszentrum J\"{u}lich and JARA, 52425 J\"{u}lich, Germany}
\author{Yuriy Mokrousov}
\email{y.mokrousov@fz-juelich.de}
\affiliation{Peter Grunberg Institut and Institute for Advanced Simulation, 
Forschungszentrum J\"{u}lich and JARA, 52425 J\"{u}lich, Germany}
\affiliation{Institute of Physics, Johannes Gutenberg-University Mainz, 55128 Mainz, Germany}

\date{\today}                                         
\maketitle

\section{Introduction} 

The anomalous Hall effect (AHE) plays a crucial role in modern condensed matter physics and material science research, as it often yields direct insight into fundamental physical properties~\cite{Nagaosa-2010, Sinova-2015, Hall-1879}. In absence of  noncollinear magnetic order, the behavior of the Hall resistivity in an external magnetic field $H$ is expressed as $\rho_{xy}=R_{0}H+R_{s}M$, where $R_{0}$ and $R_{s}$ are ordinary and extraordinary Hall coefficients, respectively, and $M$ is the magnetization of the sample. Second term in the above equation is known as anomalous Hall contribution. The additional hump-like anomalies appearing on  top of the expected behavior of Hall resistivity with increasing magnetic field are often interpreted as a signature of the formation of non-collinear magnetic structures, including topological chiral particles such as magnetic skyrmions, contributing to the Hall signal via the so-called topological Hall effect. 
The latter phenomenon, perceived as one of the manifestations of chirality-sensitive Hall effect of textures~\cite{PhysRevLett.124.096602,PhysRevB.102.184407,Kipp}, has gained considerable attention in the past years as it allows for probing the existence of chiral spin textures by purely electrical means~\cite{Back_2020,GOBEL2020}. 
In this context,  after the prediction of interface-stabilized skyrmion formation in SrRuO$_3$/SrIrO$_3$ heterostructures\cite{Tokura-SrRuO3-SrIrO3-2016, Yang-SrRuO3-SrIrO3-2019}, thin films of SrRuO$_3$ (SRO) $-$ a material which is historically of great general interest in the context of spintronics applications~\cite{Chen-2013, Fang-2003, Spintronics-2004} $-$ have drawn an immense  activity in the past years aimed at the observation of skyrmion phase by the means of magneto-transport. 

 In particular, additional hump like anomalies on top of the expected AHE signal were reported recently in ultrathin films of ferromagnetic SrRuO$_3$\cite{Sohn-korea-2018, Youdi-china-2019,Qin-singapore-2019, Zhang-2020}. 
It was postulated that RuO$_6$ octahedra-titling induced by local orthorhombic-to-tetragonal structural phase transition at SrRuO$_3$/SrTiO$_3$ interface breaks structural inversion symmetry and, mediated by symmetry lowering and intrinsic strong spin-orbit coupling of $4d$ SRO states, leads to a finite Dzyaloshinskii-Moriya interaction and stabilization of the skyrmion phase\cite{Youdi-china-2019}. However, this picture of skyrmion phase stabilization in the system was questioned by several works\cite{Groenendijk-2020,Lena-2020,Daisuke-2018, Zeise-2019,Kimbell-2020}.
Very recently, the behavior of the Hall signal measured in SrRuO$_3$ films deposited on Pr$_{0.7}$Ca$_{0.3}$MnO$_{3}$ (PCMO)\cite{Zeise-2019} $-$ reminiscent of that driven the topological Hall effect $-$ was interpreted as arising from a superposition of Hall effect contributions originated separately in tetragonal and orthorhombic SrRuO$_3$ layers. Depending on strain\cite{Daisuke-2011-strain,Daisuke-2013-tetra,MAE-Tokura-2004} or thickness of PCMO\cite{Zeise-2019}, modifications in the RuO$_6$ octahedral tilting were reported for ultrathin ferromagnetic SRO layers, while for larger strain or for thicker PCMO films, crystal symmetry of deposited ultrathin SrRuO$_3$ was found to undergo an orthorhombic-to-tetragonal transition associated with a large  change in the RuO$_6$ octahedral tilting angle. Notably, an opposite sign of the AHE for tetragonal and orthorhombic SrRuO$_3$ films was reported, and it was speculated that this was a probable reason for an observed additional peak in the field dependence of the AHE\cite{Kimbell-2020,Zeise-2019}. 

The existing controversy motivated us to study the AHE of ferromagnetic SRO films from first principles theory in order to gain a better understanding of the AHE in this system in relation to its structural properties.   
In this work we investigate the electronic and Hall transport properties of ferromagnetic SRO trilayer by tailoring the oxygen-octahedra tilting angle which is driven by the substrate induced compressive strain. 
By performing first principles density functional theory (DFT) calculations we find ferromagnetic SRO films grown on SrTiO$_3$ (corresponding to a strain of $-$0.47$\%$) to have an orthorhombic (both tilting and rotation) distorted structure, and with an increasing amount of substrate induced compressive strain the octahedral tilting angle is found to be suppressed gradually, with  ferromagnetic SRO films grown on NdGaO$_3$ (NGO) (corresponding to a strain of $-$1.7$\%$) found to be stabilized in the tetragonal distorted structure (with zero tilting) with magnetic moments pointing out of the plane of the film in complete agreement with experimental observations\cite{Daisuke-2013-tetra,MAE-Tokura-2004,PRL-SRO2020,Xia-PRB2009,Toyota-MIT2005}. 

Our  calculations of the AHE predict a positive value of the anomalous Hall conductivity ($+$76\,S/cm) at $-$1.7$\%$ strain, whereas the anomalous Hall conductivity is found to be negative ($-$156\,S/cm) at $-$0.47$\%$ strain, which is in good quantitative agreement with experimental data\cite{Fang-2003, Zeise-2019, Sohn-ARPES-2019, Zeise-Hall-effect-2013}. We attribute the found behavior of the AHE to the nodal point dynamics in the electronic structure arising in response to tailoring the oxygen octahedral distortion driven by the substrate induced strain. 
We also estimated the AHE as a function of reduced magnetization to mimic the experimentally observed temperature dependence of the AHE in the system, achieving a good agreement with the experimentally observed trend\cite{Sohn-ARPES-2019,Zeise-2019}. We believe that our findings contribute significantly to understanding the physics of the anomalous Hall effect in this delicate material, and point towards new routes of engineering the anomalous Hall effect properties in complex oxides.


\section{Computational Details}

DFT calculations were carried out with two different approaches: the full-potential linearized augmented plane wave (FLAPW) method as implemented in the J\"{u}lich DFT code FLEUR \cite{fleur}, and the plane-wave projected augmented wave (PAW) method as implemented in Vienna ab initio Simulation Package (VASP)\cite{vasp,paw}.
The structural optimization of the bulk, as well as thin film structures, was carried out using the VASP code maintaining the symmetry of the crystal. 
The positions of the atoms were relaxed towards equilibrium until the Hellman-Feynman forces became less than 0.001\,eV/\AA. 
The Monkhorst-Pack\cite{monk} $k$-point mesh of 8$\times$8$\times$6 was used for structural optimization of bulk SRO. This choice of the $k$-mesh and a plane-wave cutoff of 500\,eV were found to provide a good convergence of the total energy.

The Monkhorst-Pack $k$-point mesh of 10$\times$10$\times$1 was used for the structural optimization of SrO-terminated SRO thin films, consisting of three unit cells of SRO along the $z$-axis, considering different strains. 
For the structural optimization of the thin film with different compressive strain in the plane wave basis, we included 20\,\AA\, of vacuum to minimize the interaction between periodically repeated images along the $z$-axis. Then we carried out the structural optimization of thin film structure by relaxing the internal positions allowing for tilting and rotation of RuO$_6$ octahedra and keeping the in-plane lattice parameters fixed at different strain values.

Using relaxed atomic positions of SRO trilayer, total energy calculations of different structures, the electronic structure calculations including the effect of spin-orbit coupling (SOC) and the AHE calculations were carried out with the film version of the FLEUR code\cite{fleur}. For self-consistent calculations with the LAPW basis set a plane-wave cutoff of $k_{max}= 4.0$\,a.u.$^{-1}$ and a $k$-mesh of 12$\times$12 in the two-dimensional Brillouin zone were found to be sufficient for the convergence of the total energy. The plane wave cutoff for the potential (g$_{max}$) and exchange-correlation potential (g$_{max,xc}$) were set to 12.6 and 10.5\,a.u.$^{-1}$, respectively. 
The muffin-tin radii for Sr, Ru, O were set to 2.80\, a.u., 2.32\,a.u., and 1.31\,a.u., respectively. For calculations of the magnetic anisotropy energy, the effect of SOC was included self-consistently using 24$\times$24 $k$-points in the two-dimensional Brillouin zone. 
We used the Vosko-Wilk-Nusair (VWN) \cite{VWN} exchange-correlation functional within the local density approximation (LDA) for self-consistent calculations.

\begin{figure}[t!]
\begin{center}
\rotatebox{0}{\includegraphics[width=0.354\textwidth]{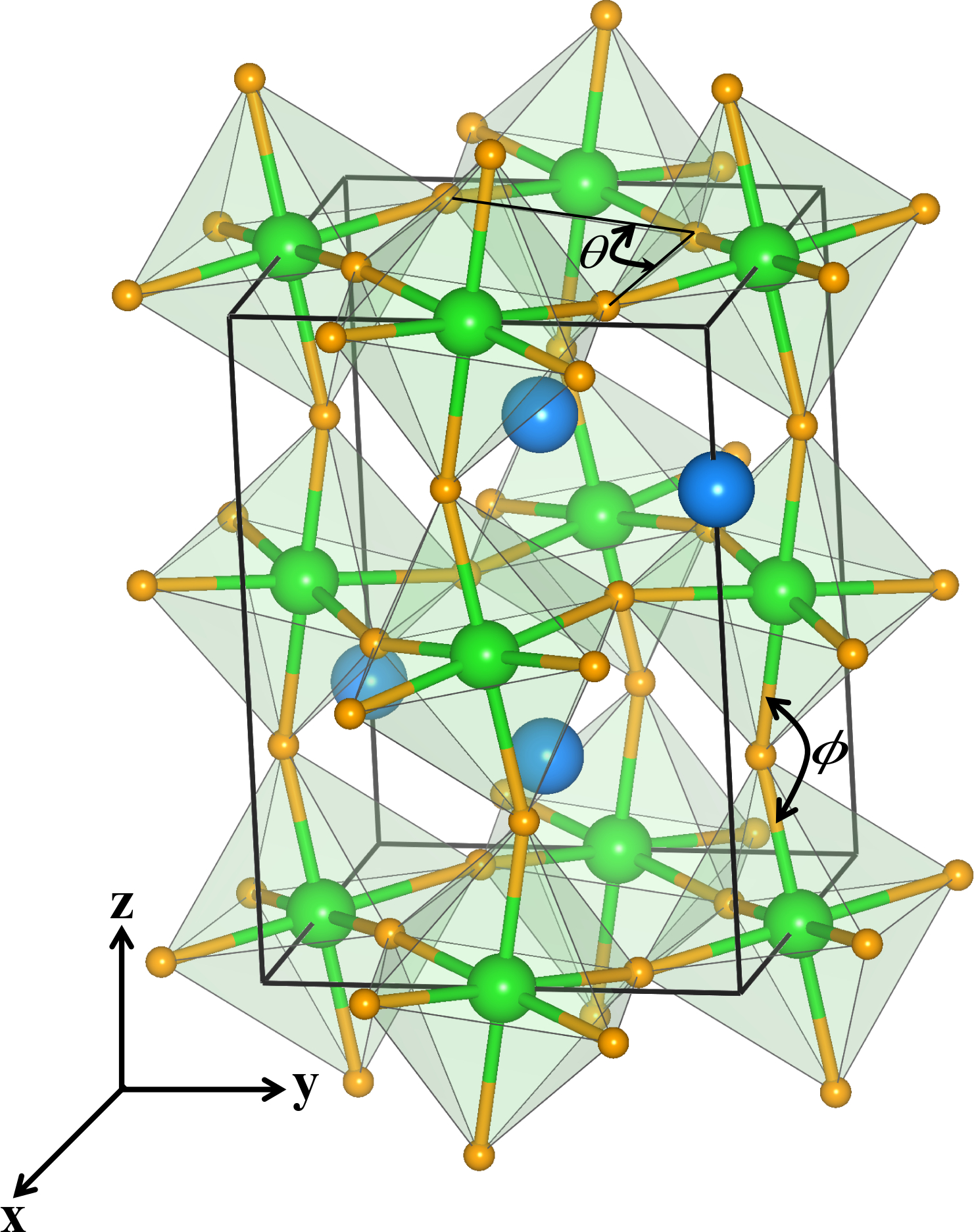}}
\end{center}
\caption{Left panel: Orthorhombic crystal structure of bulk SRO, with green-, blue- and yellow-colored spheres representing Ru, Sr, and O atoms, respectively. Tilting angle, ($180^{0}-\phi$)/2, and rotation angle, ($90^{0}-\theta$)/2, of oxygen octahedra are  marked. 
}
\label{figure:1}
\end{figure}

\section{Structural properties}
Bulk SRO is stabilized in an orthorhombic crystal structure below 850\,K with GdFeO$_3$-type distortion~\cite{Koster-2012} characterized by the tilting of the RuO$_6$ octahedra in alternate directions away from the $z$-axis and the rotation of the octahedra, as shown in Fig. 1. It is a ferromagnetic metal with Curie temperature of 160 K and magnetic moment of 1.1$-$1.7 $\mu_{B}$/f.u.\cite{SciRep-2017, James-2008}. 
The optimized lattice parameters of the bulk SRO, obtained while keeping the symmetry of the structure fixed were found to be in good agreement with previous studies \cite{Mahadevan-2009, SciRep-2017, James-2008}. We find the distortion of RuO$_6$ octahedra which manifests in unequal bond lengths and deviations of O-Ru-O bond angles away from  90$^{\circ}$:  the optimized tilting angle, $(180-\phi)/2$, is found to be 10.56$^\circ$ (corresponding to a Ru-O-Ru angle of 159$^\circ$), and rotation angle, $(90-\theta)/2$, is found to be 7.56$^{\circ}$.  

\begin{figure*}[ht!]
\begin{center}
\rotatebox{0}{\includegraphics[width=0.9\textwidth]{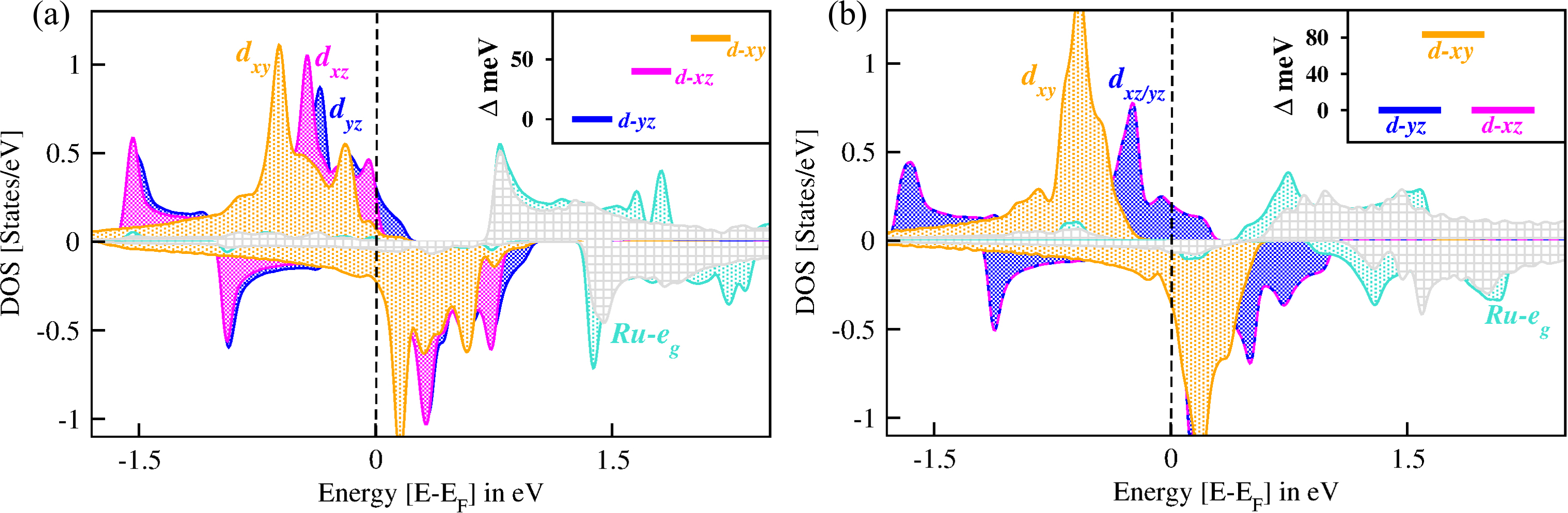}}
\end{center}
\caption{(a) Spin-polarized density of states of SRO trilayer at $-$0.47$\%$ strain as computed in LDA, projected onto the octahedral crystal-field split Ru-t$_{2g}$ states. An orthorhombic distortion of RuO$_6$ octahedra gives rise to a large splitting of the Ru-t$_{2g}$ states. (b) Spin-polarized density of states of SRO trilayer at $-$1.7$\%$ strain. A tetragonal distortion of RuO$_6$ octahedra keeps the Ru-$d_{xz/yz}$ states degenerate.
The insets in (a) and (b) show the energy level splitting, $\Delta(\epsilon_{yz/xz}-\epsilon_{xy})$, of Ru-t$_{2g}$ states.
}
\label{figure:3}
\end{figure*}
 
To simulate the thin film structure with different compressive strain, we consider a free standing tri-layer SrO-terminated SrRuO$_3$ film grown on different substrate lattice parameters. We take the optimized lattice constant of bulk SRO as the reference value (i.e. having zero strain) and define the compressive strain value with respect to it. Starting from the 
 SRO trilayer film, grown on the SrTiO$_3$ (STO) substrate (corresponding to $-$0.47$\%$ of compressive strain), strain is increased gradually up to the value of $-$1.7$\%$, 
 which corresponds to SRO thin films grown on the NdGaO$_3$ (NGO) substrate. 
For each value of strain, we first optimize the atomic positions, while keeping the symmetry of the structure fixed. 
When doing so, we observe a gradual suppression of RuO$_6$ octahedra's tilting angle compared to that of the $-$0.47$\%$-strain situation, with the trilayer in which  RuO$_6$ tilting at $-$1.7$\%$ of strain is completely suppressed found to be energetically most stable.
In the remaining part of the discussion, we focus only on two values of compressive strain: $-$0.47$\%$ and $-$1.7$\%$, which correspond to SRO thin films grown on SrTiO$_3$ and NdGaO$_3$ substrate, respectively. 


Our calculations indicate that at the strain of $-$0.47$\%$ the SRO trilayer with orthorhombic distortion of RuO$_6$ octahedra (i.e.~with simultaneous tilt and rotation)  is by 48\,meV/Ru lower than the SRO film with  tetragonal distortion (i.e.~only with octahedral rotation), while the latter structure is by 25 meV/Ru lower in energy than the orthorhombically  distorted structure at $-$1.7$\%$ strain in agreement with the experimental observations~\cite{Daisuke-2011-strain,Daisuke-2013-tetra,MAE-Tokura-2004}. For the calculations presented below we thus assume that at  $-$0.47$\%$ strain the SRO trilayer has orthorhombic RuO$_6$ octahedral distortion (i.e.~having an orthorhombic structure),
whereas it has a tetragonal distortion of RuO$_6$ at $-$1.7$\%$ strain 
(i.e.~having a tetragonal structure). The changes in the octahedral distortion in terms of the compression of bond length along the $z$-axis and tilting angle for the strain values of $-$0.47$\%$ and $-$1.7$\%$ have a crucial impact on the energetic position of Ru-t$_{2g}$ states, thus directly influencing the electronic structure for two cases, as discussed later.

\section{Electronic structure of SRO films}
For both cases of  strain, i.e.~at $-$0.47$\%$ and $-$1.7$\%$, the ferromagnetic (FM) state was found to be more stable as compared to the anti-ferromagnetic or non-magnetic one.
Assuming the FM spin structure, in Fig. 2(a) and (b) we show the orbitally-resolved density of states (DOS) as calculated with LDA, for the SRO trilayer considering the strain of $-$0.47$\%$ and $-$1.7$\%$, respectively. For both  cases, states close to the Fermi level are dominated by the Ru-t$_{2g}$ states, 
which exhibit a finite hybridization with the O-$p$ states predominantly in the lower region of the valence band (not  shown in the plots).

We observe a clear difference in the DOS between the two strain cases. Due to the orthorhombic distortion of RuO$_6$ at $-$0.47$\%$ stain, the degeneracy of the Ru-$t_{2g}$ is lifted, while for the case of $-$1.7$\%$ strain, due to complete suppression of octahedral tilting (tetragonal distortion), the degeneracy is lifted only partially, with two  degenerate Ru-$d_{xz/yz}$ bands shifted in energy with respect to a single Ru-$d_{xy}$ band. Due to the octahedral elongation along the $z$-axis with the application of compressive strain,  $d_{xz/yz}$ states are stabilized by being pushed down in energy leading in turn to the reordering of Ru-t$_{2g}$ states with $d_{xz/yz}$ states followed by the $d_{xy}$ state.

To quantify the energy level splitting of  Ru-t$_{2g}$ bands, we compute the energy-level diagram of Ru-$d$ states, employing the technique of maximally localized Wannier functions (MLWFs)\cite{Marzari-2012, Mostofi-2014,Hanke-2015,Freimuth-2008}, considering only the Ru-$d$ Hamiltonian constructed out of non-spin-polarized LDA calculations. With the information on the energy level position of Ru-$d$ states, obtained from the real-space representation of the Hamiltonian in the MLWFs basis, we calculate the energy level difference between three Ru-t$_{2g}$ states. In the inset of  Fig. 2(a) and (b) we show the energy level splitting of Ru-t$_{2g}$ states for the strain of $-$0.47$\%$ and $-$1.7$\%$, respectively. At $-$1.7$\%$ of strain, driven by the tetragonal distortion, Ru-$d_{xz/yz}$ states become  degenerate with an energy splitting of 80\,meV  between $d_{xz/yz}$ and $d_{xy}$ orbitals. On the other hand, at $-$0.47$\%$ of strain, all three t$_{2g}$ states are split off by an almost equal amount of 30 meV.

\begin{figure*}[ht!]
\begin{center}
\rotatebox{0}{\includegraphics[width=0.98\textwidth]{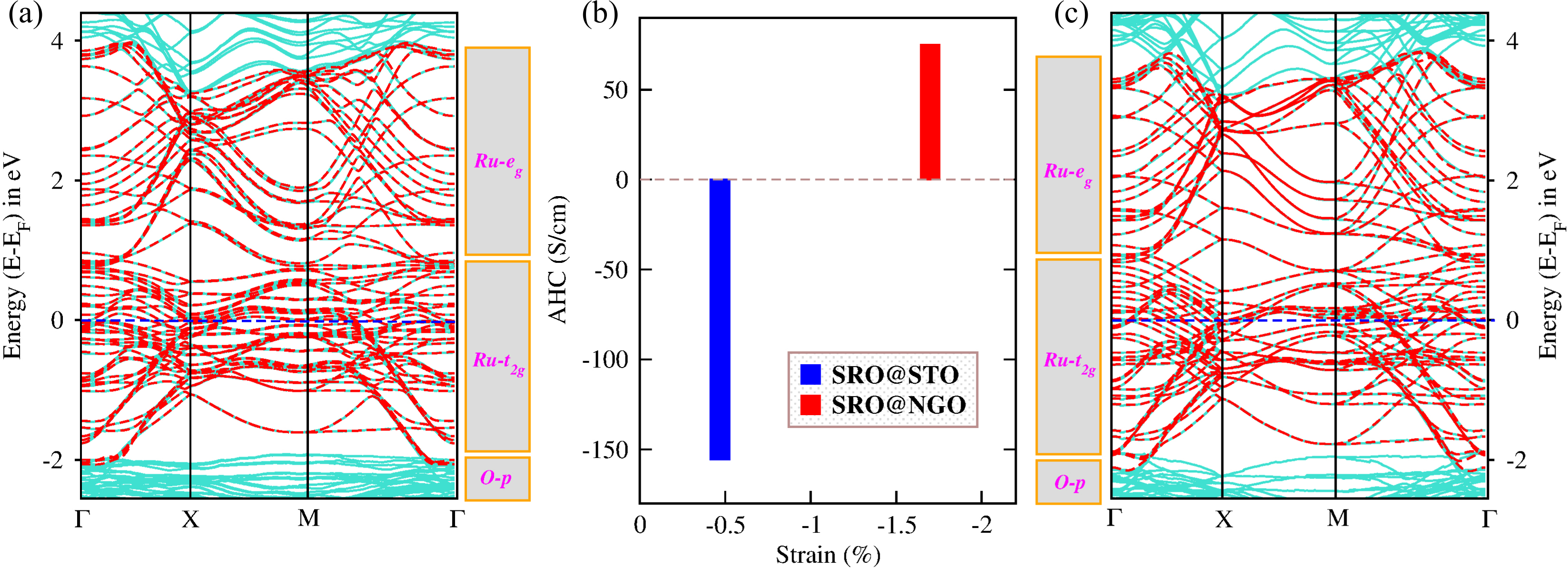}}
\end{center}
\caption{ (a), (c) The band structure of SRO trilayer grown on STO ($-$0.47$\%$ strain, a) and NGO ($-$1.7$\%$ strain, c), respectively, in the ferromagnetic  state with  magnetization along the $z$-axis, including the effect of SOC. Gray lines: GGA+SOC first principles electronic bands. Red lines: Wannier-interpolated band structure. 
The dominant orbital character of the states is shown on the side.
(b) Corresponding computed anomalous Hall conductivity (AHC). 
The blue bar shows the negative AHC of SRO trilayer grown on STO, and the red bar shows the positive value of the AHC of SRO trilayer grown on NGO.}
\label{figure:4}
\end{figure*}

As also evidenced in Fig. 2(a) and (b), owing to  the orthorhombic distortion of oxygen octahedra, the bandwidth is slightly narrower in the SRO film at $-$0.47$\%$ strain as compared to that at $-$1.7$\%$ strain. This suggests an increase in itinerancy of the SRO film at $-$1.7$\%$ strain, which is  consistent with the experimental observation of increased Curie temperature (155 K) compared to that of SRO films at $-$0.47$\%$ strain (150K). Within the mean-field theory approach, the Curie temperature, T$_c$, is proportional to $\Delta$, 
T$_c\sim\Delta$
, where $\Delta$ is the energy difference between the ferromagnetic and paramagnetic ground state. Using the mean-field theory approach, we estimate the ratio of the T$_c$ of SRO films at $-$0.47$\%$ and $-$1.7$\%$ (1.06),  finding it to be in a close agreement with the experimentally measured value (1.03), indicating that the description of the magnetic ground state within LDA is quite accurate.

\section{Magnetocrystalline anisotropy energy}
Next, we investigate the magnetic anisotropy energy (MAE) of the trilayer SRO films at $-$0.47$\%$ and $-$1.7$\%$ compressive strain. We obtain the MAE from the difference in the total energy of the ferromagnetic state with spins aligned along the crystal axes $x$ ($y$), with the total energy of the system when the magnetization is pointing along the $z$-axis. 
For the case of compressive strain studied here, irrespective of its magnitude, the out-of-plane direction is found to be the easy axis of the system in complete agreement with experimental observations\cite{Daisuke-2013-tetra, MAE-Tokura-2004}. 
Magnetic anisotropy energy of the trilayer SRO films at $-$0.47$\%$ strain is found to be $-$0.455 meV/Ru which is in a qualitative agreement with the experimentally measured value~\cite{MAE-Ziese-2010}. At $-$1.7$\%$ strain, the calculated value of the MAE constitutes about $-$0.67 meV/Ru. The calculated  averaged spin moment at the Ru ion is found to be 1.20 (1.14)\,$\mu_B$ at $-$0.47$\%$ ($-$1.70$\%$) strain, which is consistent with the low spin state of Ru$^{4+}$ ion ($d^{4}$:t$^{3}_{2g \uparrow}$,t$^{1}_{2g \downarrow}$). The orbital moments of Ru$^{2+}$($d^{4}$) point in the same direction as the spin moments, which is expected due to more than half-filled Ru-t$_{2g}$ sub-shell.

\section{Anomalous Hall conductivity}

Having understood the electronic structure and the magneto-crystalline anisotropy energy of the SRO trilayer films at $-$0.47$\%$ and $-$1.7$\%$ strain, next we proceed to investigate the anomalous Hall effect of the system, motivated in part by the recently observed anomalies in the behavior of the AHE as a function of an applied magnetic field in this material~\cite{Youdi-china-2019, Sohn-korea-2018,Zeise-2019}
Here, we assess the intrinsic Berry curvature contribution to the AHE employing the Wannier interpolation technique\cite{Wang-Souza}. To compute the Berry curvature, we first construct a tight-binding MLWFs Hamiltonian 
projected from the LDA+SOC Bloch wave-functions. Atomic-orbital-like MLWFs of Ru-t$_{2g}$ and -e$_{g}$ states were used to construct the minimal tight-binding Hamiltonian, which reproduces the spectrum of the system in a wide energy window around the Fermi energy. 
In Fig. 3(a) and 3(c), we show the comparison of the {\it ab initio} LDA+SOC band structure of SRO trilayer at $-$0.47$\%$ and $-$1.7$\%$ strain, respectively, with that obtained by diagonalization of Ru-$d$ projected Wannier Hamiltonian, finding an excellent agreement in the region of $\pm3$\,eV with respect to the Fermi energy for both strain cases.
In the latter plots, one can see a clear difference in the band structure along the $\Gamma$X direction in the Brillioun Zone for the SRO trilayer at $-$0.47$\%$ and  $-$1.7$\%$ strain. Due to the suppression of the oxygen octahedral tilting at $-$1.7$\%$ strain, the bands remain degenerate along  $\Gamma$X, whereas for $-$0.47$\%$ strain due to strong oxygen octahedral distortion 
the degeneracy is lifted. For the case of  $-$1.7$\%$ strain we find that the number of majority and minority bands crossing the Fermi energy is larger than that of the SRO film at $-$0.47$\%$ strain.

\begin{figure*}[ht!]
\begin{center}
\rotatebox{0}{\includegraphics [width=0.95\textwidth]{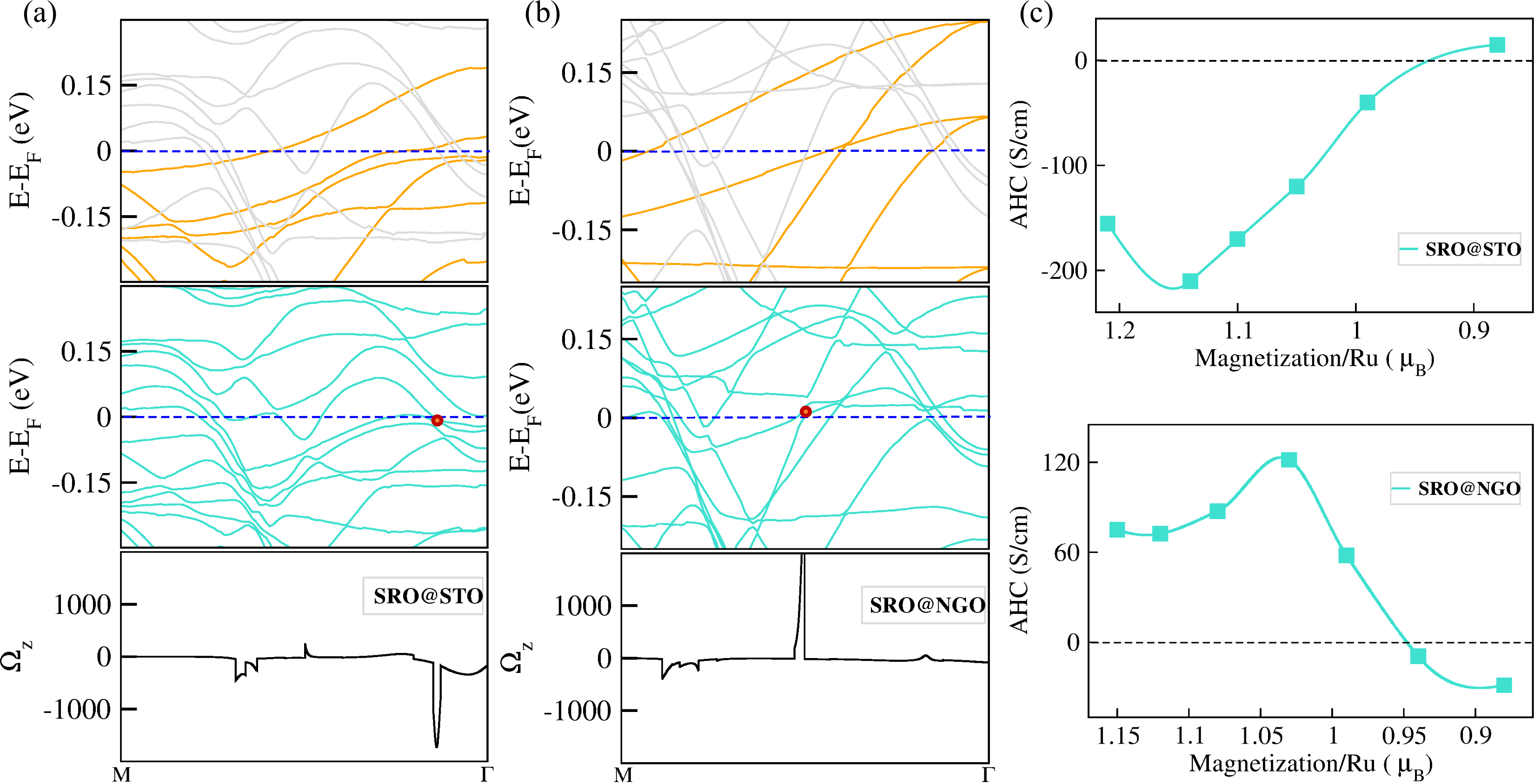}}
\end{center}
\caption{ (a-b): Upper panel: Band structures for SRO trilayer grown on STO ($-$0.47$\%$ strain, a) and NGO ($-$1.7$\%$ strain, b), respectively, near Fermi energy along the M$\Gamma$ high symmetry direction, in the absence of SOC. Orange lines represent majority bands and grey lines minority bands. Crossings of  majority and minority bands in the absence of SOC lead to the emergence of nodal points. Middle panel: electronic band structure near the Fermi level in the presence of SOC. SOC opens a gap around the nodal points marked with red dots. Lower panel: corresponding distribution of the Berry curvature of the occupied states. Large peaks in the Berry curvature can be observed near the marked nodal points. (c) Computed AHC as a function of the size of the magnetization of SRO trilayer at $-$0.47$\%$ (upper panel) and $-$1.7$\%$ (lower panel) strain. The change in the magnetization is realized by tuning the magnitude of exchange splitting inside Ru ions. }
\label{figure:5}
\end{figure*}

From the Wannier Hamiltonian the Berry curvature was calculated according to the linear-response expression as given by 
\begin{eqnarray}
\begin{aligned}
\Omega_{n}(\mathbf{k}) = -\hslash^{2} \sum_{n \neq m} \frac{\operatorname{2 Im} \langle u_{n\mathbf{k}}| \hat{ v}_{x}|u_{m\mathbf{k}}\rangle \langle u_{n\mathbf{k}}|\hat{v}_{y}|u_{m\mathbf{k}}\rangle}{(\epsilon_{n\mathbf{k}}-\epsilon_{m\mathbf{k}})^{2}},
\end{aligned}
\end{eqnarray}
 where $\Omega_{n}(\mathbf{k})$ is the Berry curvature of band $n$, $\hslash\hat{v}_{i} ={\partial \hat{H}(\mathbf{k})}/{\partial k_{i}} $ is the $i$'th velocity operator, $u_{n\mathbf{k}}$ and $\epsilon_{n\mathbf{k}}$ are the eigenstates and eigenvalues of the Hamiltonian $\hat{H}(\mathbf{k})$, respectively.
Using the tight-binding Hamiltonian constructed from atomic orbital-like Wannier functions of Ru-$d$ states for the two strain cases, as shown in Fig. 3(a) and (c), we calculate the Berry curvature on a $50\times 50$ $k$-mesh employing an adaptive $5\times 5$ refinement scheme\cite{Yao-2004} at points where the value of the Berry curvature exceeds 50\,a.u. These numerical parameters provide well-converged values of the  anomalous Hall conductivity (AHC) determined as 
\begin{eqnarray}
\begin{aligned}
\sigma_{x y}=& -\hbar e^{2} \int_{BZ} \frac{d \bf{k}}{(2 \pi)^{2}}  \Omega(\mathbf{k}),
\end{aligned}
\end{eqnarray}
where $\Omega(\mathbf{k})$ is the sum (for each $k$-point) of Berry curvature of the occupied states and the broadening of 45\,meV was used.
We have recently shown that the AHC values obtained in such a way for SRO films are stable with respect to the choice of the MLWFs reproducing the band structure in the whole energy window of occupied states~\cite{JAP-2020}. 

Our calculations of the AHC 
of the FM SRO trilayer for two values of compressive strain are shown in Fig. 3(b). We observe that  the sign of the AHC for the SRO films at $-$0.47$\%$  of strain is negative, and it becomes positive at $-$1.7$\%$ of  strain, which is in complete agreement with the experimental observations~\cite{Fang-2003,Zeise-2019, Sohn-ARPES-2019}. To understand the impact of the octahedral tilting and rotation on the sign and strength of the AHE, we start with a tetragonal distorted (zero tilting) SRO film at $-$1.7$\%$ strain with a positive AHC and gradually introduce the octahedral tilting. With the increasing value of the octahedral tilting which mimics the SRO films at $-$0.47$\%$ strain, the computed AHC displays a gradual decrease from a positive value, and changes sign for larger tilting. This indicates that symmetry lowering realized by  octahedral distortion in terms of tilting and rotation, followed by a corresponding  redistribution of the bands plays a crucial role in shaping the  strength and sign of the AHC in strained SRO films.

To gain a better insight into the microscopic origin of the observed sign changes in the  Hall response for large compressive strain of SRO trilayer films, for which oxygen octahedral tilting is completely suppressed as compared to that of the SRO films at $-$0.47$\%$ strain, we closely investigate the interplay of the Berry curvature with the electronic band structure for the two cases, with and without SOC. The electronic band structure of the SRO trilayer at $-$0.47$\%$ and $-$1.7$\%$ strain with and without  SOC is shown in the upper and middle panel of Fig. 4(a) and (b), respectively. For the case of the trilayer at $-$1.7$\%$ strain, we find that the number of majority and minority bands crossing  the Fermi energy is larger than that of the SRO film at $-$0.47$\%$ strain. In the absence of SOC, nodal points are formed due to the crossing of the majority (orange lines) and minority (grey lines) bands at the Fermi energy, 
in  Fig. 4 (a) and (b). When SOC is switched on, the degeneracy of the nodal points is lifted as shown with red dots and 
small gaps in the electronic structure, which give rise to large peaks in the Berry curvature, are formed.
The sign and magnitude of the AHC depend on the energies of the gapped nodal points relative to the Fermi energy\cite{Sohn-ARPES-2019}. At the strain of $-$0.47$\%$, large negative contribution of the Berry curvature is induced by the gapped nodal points found below the Fermi energy close to the $\Gamma$-point, whereas at the strain of $-$1.7$\%$ gapped nodal points lie above the Fermi energy (hole-like states) inducing large positive Berry curvature. And while the total value of the AHC emerges as a result of contributions coming from various parts of the Brillouin zone, as illustrated in the case of nodal points shown in Fig.~4, the sensitive redistribution of the states with respect to the Fermi energy as well as the changes in the details of their hybridization mediated by SOC and symmetry, leading to a very large response of the AHC to applied strain in this system.


\section{ Anomalous Hall conductivity with reduced magnetization}
Finally, we address the behavior of the anomalous Hall conductivity of SRO trilayer for two compressive strain cases, with the magnitude of the magnetization, in order to make an attempt at understanding the experimentally observed temperature  dependence of the anomalous Hall conductivity in SRO thin films. Within  the physical  picture that we assume, the part of the temperature-driven variation of the AHC that we try to account for is driven by the modifications in the 
magnitude of the exchange splitting of the states and corresponding ferromagnetic magnetization $M$, rather than transversal fluctuations of the local Ru moments~\cite{Fang-2003}.  To do so, we  optimize the atomic positions of the SRO trilayer at both values of strain with reduced magnetization $M$, while keeping the symmetry of the structure fixed. 
Considering optimized structures with reduced magnetization for the two compressive strain values, we calculate the intrinsic contribution to the  AHE in the way identical to that described above.
Here also, we consider the atomic-orbital-like Wannier functions, Ru-$t_{2g}$ and e$_{g}$ to construct the tight-binding Hamiltonian, projected from the LDA+SOC Bloch wave functions for each considered magnetization value. 

In Fig. 4(c) we present the computed AHC 
as a function of reduced magnetization away from the ``equilibrium" value, obtained by scaling down the magnitude of  the exchange field inside Ru atoms, for the compressive strain of $-$0.47$\%$ and $-$1.7$\%$, respectively. At $-$0.47$\%$ strain, 
for large values of the magnetization, up to a magnitude of about 0.9\,$\mu_{B}$ the AHC remains negative, whereas at the strain of $-$1.7$\%$  it stays positive for magnetization values in a similar range. 
For  both  cases of  compressive strain, the calculated AHC, after the initial increase in the absolute value decreases sharply and changes its sign, with an overall behavior reflecting basic features of experimental findings\cite{Fang-2003,Sohn-ARPES-2019, Zeise-2019}. The found correspondence between the experimentally observed data and our calculations suggests  that in thin films of strained SRO longitudinal thermal fluctuations of Ru spin moments have a drastic impact on the AHE.

\section{Summary}
In this work, by performing first principles calculations we investigate the effect of compressive strain induced oxygen octahedral distortion in the electronic structure and anomalous Hall response of the ferromagnetic SRO ultra-thin films. We find a strong deformation of the oxygen octahedra (RuO$_6$) with an increasing amount of substrate induced compressive strain. The free standing SRO film grown on SrTiO$_3$ (in-plane strain of $-$0.47$\%$) is found to have an orthorhombic (both tilting and rotation) distorted structure, and with an increasing amount of substrate induced compressive strain the octahedral tilting angle is found to be suppressed gradually, with SRO films grown on NdGaO$_3$ (in-plane strain of $-$1.7$\%$)  stabilized in the tetragonal distorted structure (with zero tilting) in complete agreement with experimental observations\cite{Daisuke-2013-tetra,MAE-Tokura-2004}. This strong modification of the oxygen octahedral deformation with strain leads to the re-ordering of  Ru-t$_{2g}$ states which has a crucial impact on the electronic structure of SRO films, and hence on the anomalous Hall response.

From the Berry curvature calculation, we find a positive value of the anomalous Hall conductivity  of $+$76\,S/cm at $-$1.7$\%$ strain, whereas it is found to be negative ($-$156\,S/cm) at $-$0.47$\%$ strain which is in good quantitative agreement with experimental data \cite{Zeise-2019, Sohn-ARPES-2019}. We attribute the observed behavior of the anomalous Hall effect to  the nodal point dynamics in the electronic structure arising in response to tailoring the oxygen octahedral distortion driven by the substrate induced strain. We also calculate  strain-mediated anomalous Hall conductivity as a function of reduced magnetization obtained by scaling down the magnitude of  the exchange field inside Ru atoms to understanding the experimentally observed temperature  dependence of the anomalous Hall conductivity. The result is found to be in good agreement with the experimental observations \cite{Fang-2003, Zeise-2019, Sohn-ARPES-2019}, indicating a strong impact of longitudinal thermal fluctuations of Ru spin moments on the anomalous Hall effect in this system. 

Overall, our findings reveal the strong influence of the oxygen octahedra deformation, arising from the substrates induced strain, on the electronic structure and anomalous Hall response of SRO films. 
We believe that our results will help to gain a better understanding of the AHE dynamics in this fascinating system in response to its structural properties, thus further motivating the development of a spintronics paradigm whose functionality is based on engineering the  topological band structure by controlling the oxygen octahedral deformations.

{\bf Data availability.} The data that support the findings of this study are available from the corresponding author upon reasonable request.

\section{Acknowledgements}
We acknowledge extensive discussions with Ionela Lindfors-Vrejoiu, Jairo Sinova and Thomas Lorentz. We acknowledge funding from Deutsche For\-schungs\-gemeinschaft (DFG) through SPP 2137 ``Skyrmionics", the Collaborative Research Center SFB 1238, and project MO 1731/5-1. The work was funded also by the Deutsche Forschungsgemeinschaft (DFG, German Research Foundation) $-$ TRR 173 $-$ 268565370 (project A11), TRR 288 $-$ 422213477 (project B06). Y.M and S.B.\ acknowledge the DARPA TEE program through grant MIPR\# HR0011831554 from DOI.   Simulations were performed with computing resources granted by JARA-HPC  from RWTH Aachen University and Forschungszentrum J\"ulich under projects jpgi11, jiff40 and jias1f. 



\hbadness=99999 
\bibliographystyle{apsrev4-2}
\bibliography{literature}

\begin{thebibliography}{47}%
\makeatletter
\providecommand \@ifxundefined [1]{%
 \@ifx{#1\undefined}
}%
\providecommand \@ifnum [1]{%
 \ifnum #1\expandafter \@firstoftwo
 \else \expandafter \@secondoftwo
 \fi
}%
\providecommand \@ifx [1]{%
 \ifx #1\expandafter \@firstoftwo
 \else \expandafter \@secondoftwo
 \fi
}%
\providecommand \natexlab [1]{#1}%
\providecommand \enquote  [1]{``#1''}%
\providecommand \bibnamefont  [1]{#1}%
\providecommand \bibfnamefont [1]{#1}%
\providecommand \citenamefont [1]{#1}%
\providecommand \href@noop [0]{\@secondoftwo}%
\providecommand \href [0]{\begingroup \@sanitize@url \@href}%
\providecommand \@href[1]{\@@startlink{#1}\@@href}%
\providecommand \@@href[1]{\endgroup#1\@@endlink}%
\providecommand \@sanitize@url [0]{\catcode `\\12\catcode `\$12\catcode
  `\&12\catcode `\#12\catcode `\^12\catcode `\_12\catcode `\%12\relax}%
\providecommand \@@startlink[1]{}%
\providecommand \@@endlink[0]{}%
\providecommand \url  [0]{\begingroup\@sanitize@url \@url }%
\providecommand \@url [1]{\endgroup\@href {#1}{\urlprefix }}%
\providecommand \urlprefix  [0]{URL }%
\providecommand \Eprint [0]{\href }%
\providecommand \doibase [0]{https://doi.org/}%
\providecommand \selectlanguage [0]{\@gobble}%
\providecommand \bibinfo  [0]{\@secondoftwo}%
\providecommand \bibfield  [0]{\@secondoftwo}%
\providecommand \translation [1]{[#1]}%
\providecommand \BibitemOpen [0]{}%
\providecommand \bibitemStop [0]{}%
\providecommand \bibitemNoStop [0]{.\EOS\space}%
\providecommand \EOS [0]{\spacefactor3000\relax}%
\providecommand \BibitemShut  [1]{\csname bibitem#1\endcsname}%
\let\auto@bib@innerbib\@empty
\bibitem [{\citenamefont {Nagaosa}\ \emph {et~al.}(2010)\citenamefont
  {Nagaosa}, \citenamefont {Sinova}, \citenamefont {Onoda}, \citenamefont
  {MacDonald},\ and\ \citenamefont {Ong}}]{Nagaosa-2010}%
  \BibitemOpen
  \bibfield  {author} {\bibinfo {author} {\bibfnamefont {N.}~\bibnamefont
  {Nagaosa}}, \bibinfo {author} {\bibfnamefont {J.}~\bibnamefont {Sinova}},
  \bibinfo {author} {\bibfnamefont {S.}~\bibnamefont {Onoda}}, \bibinfo
  {author} {\bibfnamefont {A.~H.}\ \bibnamefont {MacDonald}},\ and\ \bibinfo
  {author} {\bibfnamefont {N.~P.}\ \bibnamefont {Ong}},\ }\href
  {https://doi.org/10.1103/RevModPhys.82.1539} {\bibfield  {journal} {\bibinfo
  {journal} {Rev. Mod. Phys.}\ }\textbf {\bibinfo {volume} {82}},\ \bibinfo
  {pages} {1539} (\bibinfo {year} {2010})}\BibitemShut {NoStop}%
\bibitem [{\citenamefont {Sinova}\ \emph {et~al.}(2015)\citenamefont {Sinova},
  \citenamefont {Valenzuela}, \citenamefont {Wunderlich}, \citenamefont
  {Back},\ and\ \citenamefont {Jungwirth}}]{Sinova-2015}%
  \BibitemOpen
  \bibfield  {author} {\bibinfo {author} {\bibfnamefont {J.}~\bibnamefont
  {Sinova}}, \bibinfo {author} {\bibfnamefont {S.~O.}\ \bibnamefont
  {Valenzuela}}, \bibinfo {author} {\bibfnamefont {J.}~\bibnamefont
  {Wunderlich}}, \bibinfo {author} {\bibfnamefont {C.~H.}\ \bibnamefont
  {Back}},\ and\ \bibinfo {author} {\bibfnamefont {T.}~\bibnamefont
  {Jungwirth}},\ }\href {https://doi.org/10.1103/RevModPhys.87.1213} {\bibfield
   {journal} {\bibinfo  {journal} {Rev. Mod. Phys.}\ }\textbf {\bibinfo
  {volume} {87}},\ \bibinfo {pages} {1213} (\bibinfo {year}
  {2015})}\BibitemShut {NoStop}%
\bibitem [{\citenamefont {Hall}(1879)}]{Hall-1879}%
  \BibitemOpen
  \bibfield  {author} {\bibinfo {author} {\bibfnamefont {E.~H.}\ \bibnamefont
  {Hall}},\ }\href {http://www.jstor.org/stable/2369245} {\bibfield  {journal}
  {\bibinfo  {journal} {American Journal of Mathematics}\ }\textbf {\bibinfo
  {volume} {2}},\ \bibinfo {pages} {287} (\bibinfo {year} {1879})}\BibitemShut
  {NoStop}%
\bibitem [{\citenamefont {Lux}\ \emph {et~al.}(2020)\citenamefont {Lux},
  \citenamefont {Freimuth}, \citenamefont {Bl\"ugel},\ and\ \citenamefont
  {Mokrousov}}]{PhysRevLett.124.096602}%
  \BibitemOpen
  \bibfield  {author} {\bibinfo {author} {\bibfnamefont {F.~R.}\ \bibnamefont
  {Lux}}, \bibinfo {author} {\bibfnamefont {F.}~\bibnamefont {Freimuth}},
  \bibinfo {author} {\bibfnamefont {S.}~\bibnamefont {Bl\"ugel}},\ and\
  \bibinfo {author} {\bibfnamefont {Y.}~\bibnamefont {Mokrousov}},\ }\href
  {https://doi.org/10.1103/PhysRevLett.124.096602} {\bibfield  {journal}
  {\bibinfo  {journal} {Phys. Rev. Lett.}\ }\textbf {\bibinfo {volume} {124}},\
  \bibinfo {pages} {096602} (\bibinfo {year} {2020})}\BibitemShut {NoStop}%
\bibitem [{\citenamefont {Redies}\ \emph {et~al.}(2020)\citenamefont {Redies},
  \citenamefont {Lux}, \citenamefont {Hanke}, \citenamefont {Buhl},
  \citenamefont {Bl\"ugel},\ and\ \citenamefont
  {Mokrousov}}]{PhysRevB.102.184407}%
  \BibitemOpen
  \bibfield  {author} {\bibinfo {author} {\bibfnamefont {M.}~\bibnamefont
  {Redies}}, \bibinfo {author} {\bibfnamefont {F.~R.}\ \bibnamefont {Lux}},
  \bibinfo {author} {\bibfnamefont {J.-P.}\ \bibnamefont {Hanke}}, \bibinfo
  {author} {\bibfnamefont {P.~M.}\ \bibnamefont {Buhl}}, \bibinfo {author}
  {\bibfnamefont {S.}~\bibnamefont {Bl\"ugel}},\ and\ \bibinfo {author}
  {\bibfnamefont {Y.}~\bibnamefont {Mokrousov}},\ }\href
  {https://doi.org/10.1103/PhysRevB.102.184407} {\bibfield  {journal} {\bibinfo
   {journal} {Phys. Rev. B}\ }\textbf {\bibinfo {volume} {102}},\ \bibinfo
  {pages} {184407} (\bibinfo {year} {2020})}\BibitemShut {NoStop}%
\bibitem [{\citenamefont {Kipp}\ \emph {et~al.}(2020)\citenamefont {Kipp},
  \citenamefont {Samanta}, \citenamefont {Lux}, \citenamefont {Merte},
  \citenamefont {Hanke}, \citenamefont {Redies}, \citenamefont {Freimuth},
  \citenamefont {Blügel}, \citenamefont {Ležaić},\ and\ \citenamefont
  {Mokrousov}}]{Kipp}%
  \BibitemOpen
  \bibfield  {author} {\bibinfo {author} {\bibfnamefont {J.}~\bibnamefont
  {Kipp}}, \bibinfo {author} {\bibfnamefont {K.}~\bibnamefont {Samanta}},
  \bibinfo {author} {\bibfnamefont {F.~R.}\ \bibnamefont {Lux}}, \bibinfo
  {author} {\bibfnamefont {M.}~\bibnamefont {Merte}}, \bibinfo {author}
  {\bibfnamefont {J.-P.}\ \bibnamefont {Hanke}}, \bibinfo {author}
  {\bibfnamefont {M.}~\bibnamefont {Redies}}, \bibinfo {author} {\bibfnamefont
  {F.}~\bibnamefont {Freimuth}}, \bibinfo {author} {\bibfnamefont
  {S.}~\bibnamefont {Blügel}}, \bibinfo {author} {\bibfnamefont
  {M.}~\bibnamefont {Ležaić}},\ and\ \bibinfo {author} {\bibfnamefont
  {Y.}~\bibnamefont {Mokrousov}},\ }\href {http://arxiv.org/abs/2007.01529}
  {\bibfield  {journal} {\bibinfo  {journal} {arXiv:2007.01529 [cond-mat]}\ }
  (\bibinfo {year} {2020})},\ \bibinfo {note} {arXiv: 2007.01529}\BibitemShut
  {NoStop}%
\bibitem [{\citenamefont {Back}\ \emph {et~al.}(2020)\citenamefont {Back},
  \citenamefont {Cros}, \citenamefont {Ebert}, \citenamefont {Everschor-Sitte},
  \citenamefont {Fert}, \citenamefont {Garst}, \citenamefont {Ma},
  \citenamefont {Mankovsky}, \citenamefont {Monchesky}, \citenamefont
  {Mostovoy}, \citenamefont {Nagaosa}, \citenamefont {Parkin}, \citenamefont
  {Pfleiderer}, \citenamefont {Reyren}, \citenamefont {Rosch}, \citenamefont
  {Taguchi}, \citenamefont {Tokura}, \citenamefont {von Bergmann},\ and\
  \citenamefont {Zang}}]{Back_2020}%
  \BibitemOpen
  \bibfield  {author} {\bibinfo {author} {\bibfnamefont {C.}~\bibnamefont
  {Back}}, \bibinfo {author} {\bibfnamefont {V.}~\bibnamefont {Cros}}, \bibinfo
  {author} {\bibfnamefont {H.}~\bibnamefont {Ebert}}, \bibinfo {author}
  {\bibfnamefont {K.}~\bibnamefont {Everschor-Sitte}}, \bibinfo {author}
  {\bibfnamefont {A.}~\bibnamefont {Fert}}, \bibinfo {author} {\bibfnamefont
  {M.}~\bibnamefont {Garst}}, \bibinfo {author} {\bibfnamefont
  {T.}~\bibnamefont {Ma}}, \bibinfo {author} {\bibfnamefont {S.}~\bibnamefont
  {Mankovsky}}, \bibinfo {author} {\bibfnamefont {T.~L.}\ \bibnamefont
  {Monchesky}}, \bibinfo {author} {\bibfnamefont {M.}~\bibnamefont {Mostovoy}},
  \bibinfo {author} {\bibfnamefont {N.}~\bibnamefont {Nagaosa}}, \bibinfo
  {author} {\bibfnamefont {S.~S.~P.}\ \bibnamefont {Parkin}}, \bibinfo {author}
  {\bibfnamefont {C.}~\bibnamefont {Pfleiderer}}, \bibinfo {author}
  {\bibfnamefont {N.}~\bibnamefont {Reyren}}, \bibinfo {author} {\bibfnamefont
  {A.}~\bibnamefont {Rosch}}, \bibinfo {author} {\bibfnamefont
  {Y.}~\bibnamefont {Taguchi}}, \bibinfo {author} {\bibfnamefont
  {Y.}~\bibnamefont {Tokura}}, \bibinfo {author} {\bibfnamefont
  {K.}~\bibnamefont {von Bergmann}},\ and\ \bibinfo {author} {\bibfnamefont
  {J.}~\bibnamefont {Zang}},\ }\href {https://doi.org/10.1088/1361-6463/ab8418}
  {\bibfield  {journal} {\bibinfo  {journal} {Journal of Physics D: Applied
  Physics}\ }\textbf {\bibinfo {volume} {53}},\ \bibinfo {pages} {363001}
  (\bibinfo {year} {2020})}\BibitemShut {NoStop}%
\bibitem [{\citenamefont {Göbel}\ \emph {et~al.}(2020)\citenamefont {Göbel},
  \citenamefont {Mertig},\ and\ \citenamefont {Tretiakov}}]{GOBEL2020}%
  \BibitemOpen
  \bibfield  {author} {\bibinfo {author} {\bibfnamefont {B.}~\bibnamefont
  {Göbel}}, \bibinfo {author} {\bibfnamefont {I.}~\bibnamefont {Mertig}},\
  and\ \bibinfo {author} {\bibfnamefont {O.~A.}\ \bibnamefont {Tretiakov}},\
  }\bibfield  {journal} {\bibinfo  {journal} {Physics Reports}\ }\href
  {https://doi.org/https://doi.org/10.1016/j.physrep.2020.10.001}
  {https://doi.org/10.1016/j.physrep.2020.10.001} (\bibinfo {year}
  {2020})\BibitemShut {NoStop}%
\bibitem [{\citenamefont {Matsuno}\ \emph {et~al.}(2016)\citenamefont
  {Matsuno}, \citenamefont {Ogawa}, \citenamefont {Yasuda}, \citenamefont
  {Kagawa}, \citenamefont {Koshibae}, \citenamefont {Nagaosa}, \citenamefont
  {Tokura},\ and\ \citenamefont {Kawasaki}}]{Tokura-SrRuO3-SrIrO3-2016}%
  \BibitemOpen
  \bibfield  {author} {\bibinfo {author} {\bibfnamefont {J.}~\bibnamefont
  {Matsuno}}, \bibinfo {author} {\bibfnamefont {N.}~\bibnamefont {Ogawa}},
  \bibinfo {author} {\bibfnamefont {K.}~\bibnamefont {Yasuda}}, \bibinfo
  {author} {\bibfnamefont {F.}~\bibnamefont {Kagawa}}, \bibinfo {author}
  {\bibfnamefont {W.}~\bibnamefont {Koshibae}}, \bibinfo {author}
  {\bibfnamefont {N.}~\bibnamefont {Nagaosa}}, \bibinfo {author} {\bibfnamefont
  {Y.}~\bibnamefont {Tokura}},\ and\ \bibinfo {author} {\bibfnamefont
  {M.}~\bibnamefont {Kawasaki}},\ }\bibfield  {journal} {\bibinfo  {journal}
  {Science Advances}\ }\textbf {\bibinfo {volume} {2}},\ \href
  {https://doi.org/10.1126/sciadv.1600304} {10.1126/sciadv.1600304} (\bibinfo
  {year} {2016}),\ \Eprint
  {https://arxiv.org/abs/https://advances.sciencemag.org/content/2/7/e1600304.full.pdf}
  {https://advances.sciencemag.org/content/2/7/e1600304.full.pdf} \BibitemShut
  {NoStop}%
\bibitem [{\citenamefont {Meng}\ \emph {et~al.}(2019)\citenamefont {Meng},
  \citenamefont {Ahmed}, \citenamefont {Baćani}, \citenamefont {Mandru},
  \citenamefont {Zhao}, \citenamefont {Bagués}, \citenamefont {Esser},
  \citenamefont {Flores}, \citenamefont {McComb}, \citenamefont {Hug},\ and\
  \citenamefont {Yang}}]{Yang-SrRuO3-SrIrO3-2019}%
  \BibitemOpen
  \bibfield  {author} {\bibinfo {author} {\bibfnamefont {K.-Y.}\ \bibnamefont
  {Meng}}, \bibinfo {author} {\bibfnamefont {A.~S.}\ \bibnamefont {Ahmed}},
  \bibinfo {author} {\bibfnamefont {M.}~\bibnamefont {Baćani}}, \bibinfo
  {author} {\bibfnamefont {A.-O.}\ \bibnamefont {Mandru}}, \bibinfo {author}
  {\bibfnamefont {X.}~\bibnamefont {Zhao}}, \bibinfo {author} {\bibfnamefont
  {N.}~\bibnamefont {Bagués}}, \bibinfo {author} {\bibfnamefont {B.~D.}\
  \bibnamefont {Esser}}, \bibinfo {author} {\bibfnamefont {J.}~\bibnamefont
  {Flores}}, \bibinfo {author} {\bibfnamefont {D.~W.}\ \bibnamefont {McComb}},
  \bibinfo {author} {\bibfnamefont {H.~J.}\ \bibnamefont {Hug}},\ and\ \bibinfo
  {author} {\bibfnamefont {F.}~\bibnamefont {Yang}},\ }\href
  {https://doi.org/10.1021/acs.nanolett.9b00596} {\bibfield  {journal}
  {\bibinfo  {journal} {Nano Letters}\ }\textbf {\bibinfo {volume} {19}},\
  \bibinfo {pages} {3169} (\bibinfo {year} {2019})},\ \Eprint
  {https://arxiv.org/abs/https://doi.org/10.1021/acs.nanolett.9b00596}
  {https://doi.org/10.1021/acs.nanolett.9b00596} \BibitemShut {NoStop}%
\bibitem [{\citenamefont {Chen}\ \emph {et~al.}(2013)\citenamefont {Chen},
  \citenamefont {Bergman},\ and\ \citenamefont {Burkov}}]{Chen-2013}%
  \BibitemOpen
  \bibfield  {author} {\bibinfo {author} {\bibfnamefont {Y.}~\bibnamefont
  {Chen}}, \bibinfo {author} {\bibfnamefont {D.~L.}\ \bibnamefont {Bergman}},\
  and\ \bibinfo {author} {\bibfnamefont {A.~A.}\ \bibnamefont {Burkov}},\
  }\href {https://doi.org/10.1103/PhysRevB.88.125110} {\bibfield  {journal}
  {\bibinfo  {journal} {Phys. Rev. B}\ }\textbf {\bibinfo {volume} {88}},\
  \bibinfo {pages} {125110} (\bibinfo {year} {2013})}\BibitemShut {NoStop}%
\bibitem [{\citenamefont {Fang}\ \emph {et~al.}(2003)\citenamefont {Fang},
  \citenamefont {Nagaosa}, \citenamefont {Takahashi}, \citenamefont {Asamitsu},
  \citenamefont {Mathieu}, \citenamefont {Ogasawara}, \citenamefont {Yamada},
  \citenamefont {Kawasaki}, \citenamefont {Tokura},\ and\ \citenamefont
  {Terakura}}]{Fang-2003}%
  \BibitemOpen
  \bibfield  {author} {\bibinfo {author} {\bibfnamefont {Z.}~\bibnamefont
  {Fang}}, \bibinfo {author} {\bibfnamefont {N.}~\bibnamefont {Nagaosa}},
  \bibinfo {author} {\bibfnamefont {K.~S.}\ \bibnamefont {Takahashi}}, \bibinfo
  {author} {\bibfnamefont {A.}~\bibnamefont {Asamitsu}}, \bibinfo {author}
  {\bibfnamefont {R.}~\bibnamefont {Mathieu}}, \bibinfo {author} {\bibfnamefont
  {T.}~\bibnamefont {Ogasawara}}, \bibinfo {author} {\bibfnamefont
  {H.}~\bibnamefont {Yamada}}, \bibinfo {author} {\bibfnamefont
  {M.}~\bibnamefont {Kawasaki}}, \bibinfo {author} {\bibfnamefont
  {Y.}~\bibnamefont {Tokura}},\ and\ \bibinfo {author} {\bibfnamefont
  {K.}~\bibnamefont {Terakura}},\ }\href
  {https://doi.org/10.1126/science.1089408} {\bibfield  {journal} {\bibinfo
  {journal} {Science}\ }\textbf {\bibinfo {volume} {302}},\ \bibinfo {pages}
  {92} (\bibinfo {year} {2003})},\ \Eprint
  {https://arxiv.org/abs/https://science.sciencemag.org/content/302/5642/92.full.pdf}
  {https://science.sciencemag.org/content/302/5642/92.full.pdf} \BibitemShut
  {NoStop}%
\bibitem [{\citenamefont {\ifmmode \check{Z}\else
  \v{Z}\fi{}uti\ifmmode~\acute{c}\else \'{c}\fi{}}\ \emph
  {et~al.}(2004)\citenamefont {\ifmmode \check{Z}\else
  \v{Z}\fi{}uti\ifmmode~\acute{c}\else \'{c}\fi{}}, \citenamefont {Fabian},\
  and\ \citenamefont {Das~Sarma}}]{Spintronics-2004}%
  \BibitemOpen
  \bibfield  {author} {\bibinfo {author} {\bibfnamefont {I.}~\bibnamefont
  {\ifmmode \check{Z}\else \v{Z}\fi{}uti\ifmmode~\acute{c}\else \'{c}\fi{}}},
  \bibinfo {author} {\bibfnamefont {J.}~\bibnamefont {Fabian}},\ and\ \bibinfo
  {author} {\bibfnamefont {S.}~\bibnamefont {Das~Sarma}},\ }\href
  {https://doi.org/10.1103/RevModPhys.76.323} {\bibfield  {journal} {\bibinfo
  {journal} {Rev. Mod. Phys.}\ }\textbf {\bibinfo {volume} {76}},\ \bibinfo
  {pages} {323} (\bibinfo {year} {2004})}\BibitemShut {NoStop}%
\bibitem [{\citenamefont {Sohn}\ \emph {et~al.}(2018)\citenamefont {Sohn},
  \citenamefont {Kim}, \citenamefont {Park}, \citenamefont {Choi},
  \citenamefont {Moon}, \citenamefont {Choi}, \citenamefont {Choi},
  \citenamefont {Noh}, \citenamefont {Zhou}, \citenamefont {Chang},
  \citenamefont {Han},\ and\ \citenamefont {Kim}}]{Sohn-korea-2018}%
  \BibitemOpen
  \bibfield  {author} {\bibinfo {author} {\bibfnamefont {B.}~\bibnamefont
  {Sohn}}, \bibinfo {author} {\bibfnamefont {B.}~\bibnamefont {Kim}}, \bibinfo
  {author} {\bibfnamefont {S.~Y.}\ \bibnamefont {Park}}, \bibinfo {author}
  {\bibfnamefont {H.~Y.}\ \bibnamefont {Choi}}, \bibinfo {author}
  {\bibfnamefont {J.~Y.}\ \bibnamefont {Moon}}, \bibinfo {author}
  {\bibfnamefont {T.}~\bibnamefont {Choi}}, \bibinfo {author} {\bibfnamefont
  {Y.~J.}\ \bibnamefont {Choi}}, \bibinfo {author} {\bibfnamefont {T.~W.}\
  \bibnamefont {Noh}}, \bibinfo {author} {\bibfnamefont {H.}~\bibnamefont
  {Zhou}}, \bibinfo {author} {\bibfnamefont {S.~H.}\ \bibnamefont {Chang}},
  \bibinfo {author} {\bibfnamefont {J.~H.}\ \bibnamefont {Han}},\ and\ \bibinfo
  {author} {\bibfnamefont {C.}~\bibnamefont {Kim}},\ }\href
  {https://arxiv.org/abs/1912.04757} {\bibinfo {title} {Emergence of robust 2d
  skyrmions in srruo3 ultrathin film without the capping layer}} (\bibinfo
  {year} {2018}),\ \bibinfo {note} {arXiv:1810.01615}\BibitemShut {NoStop}%
\bibitem [{\citenamefont {Gu}\ \emph {et~al.}(2019)\citenamefont {Gu},
  \citenamefont {Wei}, \citenamefont {Xu}, \citenamefont {Zhang}, \citenamefont
  {Wang}, \citenamefont {Li}, \citenamefont {Saleem}, \citenamefont {Chang},
  \citenamefont {Sun}, \citenamefont {Song}, \citenamefont {Feng},
  \citenamefont {Zhong}, \citenamefont {Liu}, \citenamefont {Zhang},
  \citenamefont {Zhu},\ and\ \citenamefont {Pan}}]{Youdi-china-2019}%
  \BibitemOpen
  \bibfield  {author} {\bibinfo {author} {\bibfnamefont {Y.}~\bibnamefont
  {Gu}}, \bibinfo {author} {\bibfnamefont {Y.-W.}\ \bibnamefont {Wei}},
  \bibinfo {author} {\bibfnamefont {K.}~\bibnamefont {Xu}}, \bibinfo {author}
  {\bibfnamefont {H.}~\bibnamefont {Zhang}}, \bibinfo {author} {\bibfnamefont
  {F.}~\bibnamefont {Wang}}, \bibinfo {author} {\bibfnamefont {F.}~\bibnamefont
  {Li}}, \bibinfo {author} {\bibfnamefont {M.~S.}\ \bibnamefont {Saleem}},
  \bibinfo {author} {\bibfnamefont {C.-Z.}\ \bibnamefont {Chang}}, \bibinfo
  {author} {\bibfnamefont {J.}~\bibnamefont {Sun}}, \bibinfo {author}
  {\bibfnamefont {C.}~\bibnamefont {Song}}, \bibinfo {author} {\bibfnamefont
  {J.}~\bibnamefont {Feng}}, \bibinfo {author} {\bibfnamefont {X.}~\bibnamefont
  {Zhong}}, \bibinfo {author} {\bibfnamefont {W.}~\bibnamefont {Liu}}, \bibinfo
  {author} {\bibfnamefont {Z.}~\bibnamefont {Zhang}}, \bibinfo {author}
  {\bibfnamefont {J.}~\bibnamefont {Zhu}},\ and\ \bibinfo {author}
  {\bibfnamefont {F.}~\bibnamefont {Pan}},\ }\href
  {https://doi.org/10.1088/1361-6463/ab2fe8} {\bibfield  {journal} {\bibinfo
  {journal} {Journal of Physics D: Applied Physics}\ }\textbf {\bibinfo
  {volume} {52}},\ \bibinfo {pages} {404001} (\bibinfo {year}
  {2019})}\BibitemShut {NoStop}%
\bibitem [{\citenamefont {Qin}\ \emph {et~al.}(2019)\citenamefont {Qin},
  \citenamefont {Liu}, \citenamefont {Lin}, \citenamefont {Shu}, \citenamefont
  {Xie}, \citenamefont {Lim}, \citenamefont {Li}, \citenamefont {He},
  \citenamefont {Chow},\ and\ \citenamefont {Chen}}]{Qin-singapore-2019}%
  \BibitemOpen
  \bibfield  {author} {\bibinfo {author} {\bibfnamefont {Q.}~\bibnamefont
  {Qin}}, \bibinfo {author} {\bibfnamefont {L.}~\bibnamefont {Liu}}, \bibinfo
  {author} {\bibfnamefont {W.}~\bibnamefont {Lin}}, \bibinfo {author}
  {\bibfnamefont {X.}~\bibnamefont {Shu}}, \bibinfo {author} {\bibfnamefont
  {Q.}~\bibnamefont {Xie}}, \bibinfo {author} {\bibfnamefont {Z.}~\bibnamefont
  {Lim}}, \bibinfo {author} {\bibfnamefont {C.}~\bibnamefont {Li}}, \bibinfo
  {author} {\bibfnamefont {S.}~\bibnamefont {He}}, \bibinfo {author}
  {\bibfnamefont {G.~M.}\ \bibnamefont {Chow}},\ and\ \bibinfo {author}
  {\bibfnamefont {J.}~\bibnamefont {Chen}},\ }\href
  {https://doi.org/https://doi.org/10.1002/adma.201807008} {\bibfield
  {journal} {\bibinfo  {journal} {Advanced Materials}\ }\textbf {\bibinfo
  {volume} {31}},\ \bibinfo {pages} {1807008} (\bibinfo {year} {2019})},\
  \Eprint
  {https://arxiv.org/abs/https://onlinelibrary.wiley.com/doi/pdf/10.1002/adma.201807008}
  {https://onlinelibrary.wiley.com/doi/pdf/10.1002/adma.201807008} \BibitemShut
  {NoStop}%
\bibitem [{\citenamefont {Zhang}\ \emph {et~al.}(2020)\citenamefont {Zhang},
  \citenamefont {Das}, \citenamefont {Barts}, \citenamefont {Azhar},
  \citenamefont {Si}, \citenamefont {Held}, \citenamefont {Mostovoy},\ and\
  \citenamefont {Banerjee}}]{Zhang-2020}%
  \BibitemOpen
  \bibfield  {author} {\bibinfo {author} {\bibfnamefont {P.}~\bibnamefont
  {Zhang}}, \bibinfo {author} {\bibfnamefont {A.}~\bibnamefont {Das}}, \bibinfo
  {author} {\bibfnamefont {E.}~\bibnamefont {Barts}}, \bibinfo {author}
  {\bibfnamefont {M.}~\bibnamefont {Azhar}}, \bibinfo {author} {\bibfnamefont
  {L.}~\bibnamefont {Si}}, \bibinfo {author} {\bibfnamefont {K.}~\bibnamefont
  {Held}}, \bibinfo {author} {\bibfnamefont {M.}~\bibnamefont {Mostovoy}},\
  and\ \bibinfo {author} {\bibfnamefont {T.}~\bibnamefont {Banerjee}},\
  }\bibfield  {journal} {\bibinfo  {journal} {Physical Review Research}\
  }\textbf {\bibinfo {volume} {2}},\ \href
  {https://doi.org/10.1103/physrevresearch.2.032026}
  {10.1103/physrevresearch.2.032026} (\bibinfo {year} {2020})\BibitemShut
  {NoStop}%
\bibitem [{\citenamefont {Groenendijk}\ \emph {et~al.}(2020)\citenamefont
  {Groenendijk}, \citenamefont {Autieri}, \citenamefont {van Thiel},
  \citenamefont {Brzezicki}, \citenamefont {Hortensius}, \citenamefont
  {Afanasiev}, \citenamefont {Gauquelin}, \citenamefont {Barone}, \citenamefont
  {van~den Bos}, \citenamefont {van Aert},\ and\ \citenamefont
  {et~al.}}]{Groenendijk-2020}%
  \BibitemOpen
  \bibfield  {author} {\bibinfo {author} {\bibfnamefont {D.~J.}\ \bibnamefont
  {Groenendijk}}, \bibinfo {author} {\bibfnamefont {C.}~\bibnamefont
  {Autieri}}, \bibinfo {author} {\bibfnamefont {T.~C.}\ \bibnamefont {van
  Thiel}}, \bibinfo {author} {\bibfnamefont {W.}~\bibnamefont {Brzezicki}},
  \bibinfo {author} {\bibfnamefont {J.~R.}\ \bibnamefont {Hortensius}},
  \bibinfo {author} {\bibfnamefont {D.}~\bibnamefont {Afanasiev}}, \bibinfo
  {author} {\bibfnamefont {N.}~\bibnamefont {Gauquelin}}, \bibinfo {author}
  {\bibfnamefont {P.}~\bibnamefont {Barone}}, \bibinfo {author} {\bibfnamefont
  {K.~H.~W.}\ \bibnamefont {van~den Bos}}, \bibinfo {author} {\bibfnamefont
  {S.}~\bibnamefont {van Aert}},\ and\ \bibinfo {author} {\bibnamefont
  {et~al.}},\ }\bibfield  {journal} {\bibinfo  {journal} {Physical Review
  Research}\ }\textbf {\bibinfo {volume} {2}},\ \href
  {https://doi.org/10.1103/physrevresearch.2.023404}
  {10.1103/physrevresearch.2.023404} (\bibinfo {year} {2020})\BibitemShut
  {NoStop}%
\bibitem [{\citenamefont {Malsch}\ \emph {et~al.}(2020)\citenamefont {Malsch},
  \citenamefont {Ivaneyko}, \citenamefont {Milde}, \citenamefont {Wysocki},
  \citenamefont {Yang}, \citenamefont {van Loosdrecht}, \citenamefont
  {Lindfors-Vrejoiu},\ and\ \citenamefont {Eng}}]{Lena-2020}%
  \BibitemOpen
  \bibfield  {author} {\bibinfo {author} {\bibfnamefont {G.}~\bibnamefont
  {Malsch}}, \bibinfo {author} {\bibfnamefont {D.}~\bibnamefont {Ivaneyko}},
  \bibinfo {author} {\bibfnamefont {P.}~\bibnamefont {Milde}}, \bibinfo
  {author} {\bibfnamefont {L.}~\bibnamefont {Wysocki}}, \bibinfo {author}
  {\bibfnamefont {L.}~\bibnamefont {Yang}}, \bibinfo {author} {\bibfnamefont
  {P.~H.~M.}\ \bibnamefont {van Loosdrecht}}, \bibinfo {author} {\bibfnamefont
  {I.}~\bibnamefont {Lindfors-Vrejoiu}},\ and\ \bibinfo {author} {\bibfnamefont
  {L.~M.}\ \bibnamefont {Eng}},\ }\href
  {https://doi.org/10.1021/acsanm.9b01918} {\bibfield  {journal} {\bibinfo
  {journal} {ACS Applied Nano Materials}\ }\textbf {\bibinfo {volume} {3}},\
  \bibinfo {pages} {1182} (\bibinfo {year} {2020})},\ \Eprint
  {https://arxiv.org/abs/https://doi.org/10.1021/acsanm.9b01918}
  {https://doi.org/10.1021/acsanm.9b01918} \BibitemShut {NoStop}%
\bibitem [{\citenamefont {Kan}\ \emph {et~al.}(2018)\citenamefont {Kan},
  \citenamefont {Moriyama}, \citenamefont {Kobayashi},\ and\ \citenamefont
  {Shimakawa}}]{Daisuke-2018}%
  \BibitemOpen
  \bibfield  {author} {\bibinfo {author} {\bibfnamefont {D.}~\bibnamefont
  {Kan}}, \bibinfo {author} {\bibfnamefont {T.}~\bibnamefont {Moriyama}},
  \bibinfo {author} {\bibfnamefont {K.}~\bibnamefont {Kobayashi}},\ and\
  \bibinfo {author} {\bibfnamefont {Y.}~\bibnamefont {Shimakawa}},\ }\href
  {https://doi.org/10.1103/PhysRevB.98.180408} {\bibfield  {journal} {\bibinfo
  {journal} {Phys. Rev. B}\ }\textbf {\bibinfo {volume} {98}},\ \bibinfo
  {pages} {180408} (\bibinfo {year} {2018})}\BibitemShut {NoStop}%
\bibitem [{\citenamefont {Ziese}\ \emph {et~al.}(2019)\citenamefont {Ziese},
  \citenamefont {Jin},\ and\ \citenamefont {Lindfors-Vrejoiu}}]{Zeise-2019}%
  \BibitemOpen
  \bibfield  {author} {\bibinfo {author} {\bibfnamefont {M.}~\bibnamefont
  {Ziese}}, \bibinfo {author} {\bibfnamefont {L.}~\bibnamefont {Jin}},\ and\
  \bibinfo {author} {\bibfnamefont {I.}~\bibnamefont {Lindfors-Vrejoiu}},\
  }\href {https://doi.org/10.1088/2515-7639/ab1aef} {\bibfield  {journal}
  {\bibinfo  {journal} {Journal of Physics: Materials}\ }\textbf {\bibinfo
  {volume} {2}},\ \bibinfo {pages} {034008} (\bibinfo {year}
  {2019})}\BibitemShut {NoStop}%
\bibitem [{\citenamefont {Kimbell}\ \emph {et~al.}(2020)\citenamefont
  {Kimbell}, \citenamefont {Sass}, \citenamefont {Woltjes}, \citenamefont {Ko},
  \citenamefont {Noh}, \citenamefont {Wu},\ and\ \citenamefont
  {Robinson}}]{Kimbell-2020}%
  \BibitemOpen
  \bibfield  {author} {\bibinfo {author} {\bibfnamefont {G.}~\bibnamefont
  {Kimbell}}, \bibinfo {author} {\bibfnamefont {P.~M.}\ \bibnamefont {Sass}},
  \bibinfo {author} {\bibfnamefont {B.}~\bibnamefont {Woltjes}}, \bibinfo
  {author} {\bibfnamefont {E.~K.}\ \bibnamefont {Ko}}, \bibinfo {author}
  {\bibfnamefont {T.~W.}\ \bibnamefont {Noh}}, \bibinfo {author} {\bibfnamefont
  {W.}~\bibnamefont {Wu}},\ and\ \bibinfo {author} {\bibfnamefont {J.~W.~A.}\
  \bibnamefont {Robinson}},\ }\bibfield  {journal} {\bibinfo  {journal}
  {Physical Review Materials}\ }\textbf {\bibinfo {volume} {4}},\ \href
  {https://doi.org/10.1103/physrevmaterials.4.054414}
  {10.1103/physrevmaterials.4.054414} (\bibinfo {year} {2020})\BibitemShut
  {NoStop}%
\bibitem [{\citenamefont {Kan}\ and\ \citenamefont
  {Shimakawa}(2011)}]{Daisuke-2011-strain}%
  \BibitemOpen
  \bibfield  {author} {\bibinfo {author} {\bibfnamefont {D.}~\bibnamefont
  {Kan}}\ and\ \bibinfo {author} {\bibfnamefont {Y.}~\bibnamefont
  {Shimakawa}},\ }\href {https://doi.org/10.1021/cg201070n} {\bibfield
  {journal} {\bibinfo  {journal} {Crystal Growth \& Design}\ }\textbf {\bibinfo
  {volume} {11}},\ \bibinfo {pages} {5483} (\bibinfo {year} {2011})},\ \Eprint
  {https://arxiv.org/abs/https://doi.org/10.1021/cg201070n}
  {https://doi.org/10.1021/cg201070n} \BibitemShut {NoStop}%
\bibitem [{\citenamefont {Kan}\ \emph {et~al.}(2013)\citenamefont {Kan},
  \citenamefont {Aso}, \citenamefont {Kurata},\ and\ \citenamefont
  {Shimakawa}}]{Daisuke-2013-tetra}%
  \BibitemOpen
  \bibfield  {author} {\bibinfo {author} {\bibfnamefont {D.}~\bibnamefont
  {Kan}}, \bibinfo {author} {\bibfnamefont {R.}~\bibnamefont {Aso}}, \bibinfo
  {author} {\bibfnamefont {H.}~\bibnamefont {Kurata}},\ and\ \bibinfo {author}
  {\bibfnamefont {Y.}~\bibnamefont {Shimakawa}},\ }\href
  {https://doi.org/10.1063/1.4803869} {\bibfield  {journal} {\bibinfo
  {journal} {Journal of Applied Physics}\ }\textbf {\bibinfo {volume} {113}},\
  \bibinfo {pages} {173912} (\bibinfo {year} {2013})},\ \Eprint
  {https://arxiv.org/abs/https://doi.org/10.1063/1.4803869}
  {https://doi.org/10.1063/1.4803869} \BibitemShut {NoStop}%
\bibitem [{\citenamefont {Jung}\ \emph {et~al.}(2004)\citenamefont {Jung},
  \citenamefont {Yamada}, \citenamefont {Kawasaki},\ and\ \citenamefont
  {Tokura}}]{MAE-Tokura-2004}%
  \BibitemOpen
  \bibfield  {author} {\bibinfo {author} {\bibfnamefont {C.~U.}\ \bibnamefont
  {Jung}}, \bibinfo {author} {\bibfnamefont {H.}~\bibnamefont {Yamada}},
  \bibinfo {author} {\bibfnamefont {M.}~\bibnamefont {Kawasaki}},\ and\
  \bibinfo {author} {\bibfnamefont {Y.}~\bibnamefont {Tokura}},\ }\href
  {https://doi.org/10.1063/1.1695195} {\bibfield  {journal} {\bibinfo
  {journal} {Applied Physics Letters}\ }\textbf {\bibinfo {volume} {84}},\
  \bibinfo {pages} {2590} (\bibinfo {year} {2004})},\ \Eprint
  {https://arxiv.org/abs/https://doi.org/10.1063/1.1695195}
  {https://doi.org/10.1063/1.1695195} \BibitemShut {NoStop}%
\bibitem [{\citenamefont {Jeong}\ \emph {et~al.}(2020)\citenamefont {Jeong},
  \citenamefont {Min}, \citenamefont {Woo}, \citenamefont {Kim}, \citenamefont
  {Zhang}, \citenamefont {Cho}, \citenamefont {Son}, \citenamefont {Kim},
  \citenamefont {Han}, \citenamefont {Park}, \citenamefont {Jeong},
  \citenamefont {Ohta}, \citenamefont {Lee}, \citenamefont {Noh}, \citenamefont
  {Lee},\ and\ \citenamefont {Choi}}]{PRL-SRO2020}%
  \BibitemOpen
  \bibfield  {author} {\bibinfo {author} {\bibfnamefont {S.~G.}\ \bibnamefont
  {Jeong}}, \bibinfo {author} {\bibfnamefont {T.}~\bibnamefont {Min}}, \bibinfo
  {author} {\bibfnamefont {S.}~\bibnamefont {Woo}}, \bibinfo {author}
  {\bibfnamefont {J.}~\bibnamefont {Kim}}, \bibinfo {author} {\bibfnamefont
  {Y.-Q.}\ \bibnamefont {Zhang}}, \bibinfo {author} {\bibfnamefont {S.~W.}\
  \bibnamefont {Cho}}, \bibinfo {author} {\bibfnamefont {J.}~\bibnamefont
  {Son}}, \bibinfo {author} {\bibfnamefont {Y.-M.}\ \bibnamefont {Kim}},
  \bibinfo {author} {\bibfnamefont {J.~H.}\ \bibnamefont {Han}}, \bibinfo
  {author} {\bibfnamefont {S.}~\bibnamefont {Park}}, \bibinfo {author}
  {\bibfnamefont {H.~Y.}\ \bibnamefont {Jeong}}, \bibinfo {author}
  {\bibfnamefont {H.}~\bibnamefont {Ohta}}, \bibinfo {author} {\bibfnamefont
  {S.}~\bibnamefont {Lee}}, \bibinfo {author} {\bibfnamefont {T.~W.}\
  \bibnamefont {Noh}}, \bibinfo {author} {\bibfnamefont {J.}~\bibnamefont
  {Lee}},\ and\ \bibinfo {author} {\bibfnamefont {W.~S.}\ \bibnamefont
  {Choi}},\ }\href {https://link.aps.org/doi/10.1103/PhysRevLett.124.026401}
  {\bibfield  {journal} {\bibinfo  {journal} {Phys. Rev. Lett.}\ }\textbf
  {\bibinfo {volume} {124}},\ \bibinfo {pages} {026401} (\bibinfo {year}
  {2020})}\BibitemShut {NoStop}%
\bibitem [{\citenamefont {Xia}\ \emph {et~al.}(2009)\citenamefont {Xia},
  \citenamefont {Siemons}, \citenamefont {Koster}, \citenamefont {Beasley},\
  and\ \citenamefont {Kapitulnik}}]{Xia-PRB2009}%
  \BibitemOpen
  \bibfield  {author} {\bibinfo {author} {\bibfnamefont {J.}~\bibnamefont
  {Xia}}, \bibinfo {author} {\bibfnamefont {W.}~\bibnamefont {Siemons}},
  \bibinfo {author} {\bibfnamefont {G.}~\bibnamefont {Koster}}, \bibinfo
  {author} {\bibfnamefont {M.~R.}\ \bibnamefont {Beasley}},\ and\ \bibinfo
  {author} {\bibfnamefont {A.}~\bibnamefont {Kapitulnik}},\ }\href
  {https://link.aps.org/doi/10.1103/PhysRevB.79.140407} {\bibfield  {journal}
  {\bibinfo  {journal} {Phys. Rev. B}\ }\textbf {\bibinfo {volume} {79}},\
  \bibinfo {pages} {140407} (\bibinfo {year} {2009})}\BibitemShut {NoStop}%
\bibitem [{\citenamefont {Toyota}\ \emph {et~al.}(2005)\citenamefont {Toyota},
  \citenamefont {Ohkubo}, \citenamefont {Kumigashira}, \citenamefont {Oshima},
  \citenamefont {Ohnishi}, \citenamefont {Lippmaa}, \citenamefont {Takizawa},
  \citenamefont {Fujimori}, \citenamefont {Ono}, \citenamefont {Kawasaki},\
  and\ \citenamefont {Koinuma}}]{Toyota-MIT2005}%
  \BibitemOpen
  \bibfield  {author} {\bibinfo {author} {\bibfnamefont {D.}~\bibnamefont
  {Toyota}}, \bibinfo {author} {\bibfnamefont {I.}~\bibnamefont {Ohkubo}},
  \bibinfo {author} {\bibfnamefont {H.}~\bibnamefont {Kumigashira}}, \bibinfo
  {author} {\bibfnamefont {M.}~\bibnamefont {Oshima}}, \bibinfo {author}
  {\bibfnamefont {T.}~\bibnamefont {Ohnishi}}, \bibinfo {author} {\bibfnamefont
  {M.}~\bibnamefont {Lippmaa}}, \bibinfo {author} {\bibfnamefont
  {M.}~\bibnamefont {Takizawa}}, \bibinfo {author} {\bibfnamefont
  {A.}~\bibnamefont {Fujimori}}, \bibinfo {author} {\bibfnamefont
  {K.}~\bibnamefont {Ono}}, \bibinfo {author} {\bibfnamefont {M.}~\bibnamefont
  {Kawasaki}},\ and\ \bibinfo {author} {\bibfnamefont {H.}~\bibnamefont
  {Koinuma}},\ }\href@noop {} {\bibfield  {journal} {\bibinfo  {journal}
  {Applied Physics Letters}\ }\textbf {\bibinfo {volume} {87}},\ \bibinfo
  {pages} {162508} (\bibinfo {year} {2005})}\BibitemShut {NoStop}%
\bibitem [{\citenamefont {Sohn}\ \emph {et~al.}(2019)\citenamefont {Sohn},
  \citenamefont {Lee}, \citenamefont {Kyung}, \citenamefont {Kim},
  \citenamefont {Ryu}, \citenamefont {Oh}, \citenamefont {Kim}, \citenamefont
  {Jung}, \citenamefont {Kim}, \citenamefont {Han}, \citenamefont {Noh},
  \citenamefont {Yang},\ and\ \citenamefont {Kim}}]{Sohn-ARPES-2019}%
  \BibitemOpen
  \bibfield  {author} {\bibinfo {author} {\bibfnamefont {B.}~\bibnamefont
  {Sohn}}, \bibinfo {author} {\bibfnamefont {E.}~\bibnamefont {Lee}}, \bibinfo
  {author} {\bibfnamefont {W.}~\bibnamefont {Kyung}}, \bibinfo {author}
  {\bibfnamefont {M.}~\bibnamefont {Kim}}, \bibinfo {author} {\bibfnamefont
  {H.}~\bibnamefont {Ryu}}, \bibinfo {author} {\bibfnamefont {J.~S.}\
  \bibnamefont {Oh}}, \bibinfo {author} {\bibfnamefont {D.}~\bibnamefont
  {Kim}}, \bibinfo {author} {\bibfnamefont {J.~K.}\ \bibnamefont {Jung}},
  \bibinfo {author} {\bibfnamefont {B.}~\bibnamefont {Kim}}, \bibinfo {author}
  {\bibfnamefont {M.}~\bibnamefont {Han}}, \bibinfo {author} {\bibfnamefont
  {T.~W.}\ \bibnamefont {Noh}}, \bibinfo {author} {\bibfnamefont {B.-J.}\
  \bibnamefont {Yang}},\ and\ \bibinfo {author} {\bibfnamefont
  {C.}~\bibnamefont {Kim}},\ }\href@noop {} {\bibinfo {title} {Sign-tunable
  anomalous hall effect induced by symmetry-protected nodal structures in
  ferromagnetic perovskite oxide thin films}} (\bibinfo {year} {2019}),\
  \Eprint {https://arxiv.org/abs/1912.04757} {arXiv:1912.04757
  [cond-mat.str-el]} \BibitemShut {NoStop}%
\bibitem [{\citenamefont {Bern}\ \emph {et~al.}(2013)\citenamefont {Bern},
  \citenamefont {Ziese}, \citenamefont {Dörr}, \citenamefont {Herklotz},\ and\
  \citenamefont {Vrejoiu}}]{Zeise-Hall-effect-2013}%
  \BibitemOpen
  \bibfield  {author} {\bibinfo {author} {\bibfnamefont {F.}~\bibnamefont
  {Bern}}, \bibinfo {author} {\bibfnamefont {M.}~\bibnamefont {Ziese}},
  \bibinfo {author} {\bibfnamefont {K.}~\bibnamefont {Dörr}}, \bibinfo
  {author} {\bibfnamefont {A.}~\bibnamefont {Herklotz}},\ and\ \bibinfo
  {author} {\bibfnamefont {I.}~\bibnamefont {Vrejoiu}},\ }\href
  {https://doi.org/https://doi.org/10.1002/pssr.201206500} {\bibfield
  {journal} {\bibinfo  {journal} {physica status solidi (RRL) – Rapid
  Research Letters}\ }\textbf {\bibinfo {volume} {7}},\ \bibinfo {pages} {204}
  (\bibinfo {year} {2013})},\ \Eprint
  {https://arxiv.org/abs/https://www.onlinelibrary.wiley.com/doi/pdf/10.1002/pssr.201206500}
  {https://www.onlinelibrary.wiley.com/doi/pdf/10.1002/pssr.201206500}
  \BibitemShut {NoStop}%
\bibitem [{fle()}]{fleur}%
  \BibitemOpen
  \href@noop {} {\bibinfo {title} {www.flapw.de}}\BibitemShut {NoStop}%
\bibitem [{\citenamefont {Kresse}\ and\ \citenamefont {Joubert}(1999)}]{vasp}%
  \BibitemOpen
  \bibfield  {author} {\bibinfo {author} {\bibfnamefont {G.}~\bibnamefont
  {Kresse}}\ and\ \bibinfo {author} {\bibfnamefont {D.}~\bibnamefont
  {Joubert}},\ }\href {https://doi.org/10.1103/PhysRevB.59.1758} {\bibfield
  {journal} {\bibinfo  {journal} {Phys. Rev. B}\ }\textbf {\bibinfo {volume}
  {59}},\ \bibinfo {pages} {1758} (\bibinfo {year} {1999})}\BibitemShut
  {NoStop}%
\bibitem [{\citenamefont {Bl\"ochl}(1994)}]{paw}%
  \BibitemOpen
  \bibfield  {author} {\bibinfo {author} {\bibfnamefont {P.~E.}\ \bibnamefont
  {Bl\"ochl}},\ }\href {https://doi.org/10.1103/PhysRevB.50.17953} {\bibfield
  {journal} {\bibinfo  {journal} {Phys. Rev. B}\ }\textbf {\bibinfo {volume}
  {50}},\ \bibinfo {pages} {17953} (\bibinfo {year} {1994})}\BibitemShut
  {NoStop}%
\bibitem [{\citenamefont {Monkhorst}\ and\ \citenamefont {Pack}(1976)}]{monk}%
  \BibitemOpen
  \bibfield  {author} {\bibinfo {author} {\bibfnamefont {H.~J.}\ \bibnamefont
  {Monkhorst}}\ and\ \bibinfo {author} {\bibfnamefont {J.~D.}\ \bibnamefont
  {Pack}},\ }\href {https://doi.org/10.1103/PhysRevB.13.5188} {\bibfield
  {journal} {\bibinfo  {journal} {Phys. Rev. B}\ }\textbf {\bibinfo {volume}
  {13}},\ \bibinfo {pages} {5188} (\bibinfo {year} {1976})}\BibitemShut
  {NoStop}%
\bibitem [{\citenamefont {Vosko}\ and\ \citenamefont {Nusair}(1980)}]{VWN}%
  \BibitemOpen
  \bibfield  {author} {\bibinfo {author} {\bibfnamefont {W.~L.}\ \bibnamefont
  {Vosko}, \bibfnamefont {S.H.}}\ and\ \bibinfo {author} {\bibfnamefont
  {M.}~\bibnamefont {Nusair}},\ }\href {http://dx.doi.org/10.1139/p80-159}
  {\bibfield  {journal} {\bibinfo  {journal} {Canadian Journal of Physics}\
  }\textbf {\bibinfo {volume} {58}},\ \bibinfo {pages} {1200} (\bibinfo {year}
  {1980})}\BibitemShut {NoStop}%
\bibitem [{\citenamefont {Koster}\ \emph {et~al.}(2012)\citenamefont {Koster},
  \citenamefont {Klein}, \citenamefont {Siemons}, \citenamefont {Rijnders},
  \citenamefont {Dodge}, \citenamefont {Eom}, \citenamefont {Blank},\ and\
  \citenamefont {Beasley}}]{Koster-2012}%
  \BibitemOpen
  \bibfield  {author} {\bibinfo {author} {\bibfnamefont {G.}~\bibnamefont
  {Koster}}, \bibinfo {author} {\bibfnamefont {L.}~\bibnamefont {Klein}},
  \bibinfo {author} {\bibfnamefont {W.}~\bibnamefont {Siemons}}, \bibinfo
  {author} {\bibfnamefont {G.}~\bibnamefont {Rijnders}}, \bibinfo {author}
  {\bibfnamefont {J.~S.}\ \bibnamefont {Dodge}}, \bibinfo {author}
  {\bibfnamefont {C.-B.}\ \bibnamefont {Eom}}, \bibinfo {author} {\bibfnamefont
  {D.~H.~A.}\ \bibnamefont {Blank}},\ and\ \bibinfo {author} {\bibfnamefont
  {M.~R.}\ \bibnamefont {Beasley}},\ }\href
  {https://doi.org/10.1103/RevModPhys.84.253} {\bibfield  {journal} {\bibinfo
  {journal} {Rev. Mod. Phys.}\ }\textbf {\bibinfo {volume} {84}},\ \bibinfo
  {pages} {253} (\bibinfo {year} {2012})}\BibitemShut {NoStop}%
\bibitem [{\citenamefont {Ryee}(2017)}]{SciRep-2017}%
  \BibitemOpen
  \bibfield  {author} {\bibinfo {author} {\bibfnamefont {M.~J.}\ \bibnamefont
  {Ryee}, \bibfnamefont {Siheon~Han}},\ }\href
  {https://doi.org/10.1038/s41598-017-04044-6} {\bibfield  {journal} {\bibinfo
  {journal} {Scientific Reports}\ }\textbf {\bibinfo {volume} {7}},\ \bibinfo
  {pages} {4635} (\bibinfo {year} {2017})}\BibitemShut {NoStop}%
\bibitem [{\citenamefont {Rondinelli}\ \emph {et~al.}(2008)\citenamefont
  {Rondinelli}, \citenamefont {Caffrey}, \citenamefont {Sanvito},\ and\
  \citenamefont {Spaldin}}]{James-2008}%
  \BibitemOpen
  \bibfield  {author} {\bibinfo {author} {\bibfnamefont {J.~M.}\ \bibnamefont
  {Rondinelli}}, \bibinfo {author} {\bibfnamefont {N.~M.}\ \bibnamefont
  {Caffrey}}, \bibinfo {author} {\bibfnamefont {S.}~\bibnamefont {Sanvito}},\
  and\ \bibinfo {author} {\bibfnamefont {N.~A.}\ \bibnamefont {Spaldin}},\
  }\href {https://doi.org/10.1103/PhysRevB.78.155107} {\bibfield  {journal}
  {\bibinfo  {journal} {Phys. Rev. B}\ }\textbf {\bibinfo {volume} {78}},\
  \bibinfo {pages} {155107} (\bibinfo {year} {2008})}\BibitemShut {NoStop}%
\bibitem [{\citenamefont {Mahadevan}\ \emph {et~al.}(2009)\citenamefont
  {Mahadevan}, \citenamefont {Aryasetiawan}, \citenamefont {Janotti},\ and\
  \citenamefont {Sasaki}}]{Mahadevan-2009}%
  \BibitemOpen
  \bibfield  {author} {\bibinfo {author} {\bibfnamefont {P.}~\bibnamefont
  {Mahadevan}}, \bibinfo {author} {\bibfnamefont {F.}~\bibnamefont
  {Aryasetiawan}}, \bibinfo {author} {\bibfnamefont {A.}~\bibnamefont
  {Janotti}},\ and\ \bibinfo {author} {\bibfnamefont {T.}~\bibnamefont
  {Sasaki}},\ }\href {https://doi.org/10.1103/PhysRevB.80.035106} {\bibfield
  {journal} {\bibinfo  {journal} {Phys. Rev. B}\ }\textbf {\bibinfo {volume}
  {80}},\ \bibinfo {pages} {035106} (\bibinfo {year} {2009})}\BibitemShut
  {NoStop}%
\bibitem [{\citenamefont {Marzari}\ \emph {et~al.}(2012)\citenamefont
  {Marzari}, \citenamefont {Mostofi}, \citenamefont {Yates}, \citenamefont
  {Souza},\ and\ \citenamefont {Vanderbilt}}]{Marzari-2012}%
  \BibitemOpen
  \bibfield  {author} {\bibinfo {author} {\bibfnamefont {N.}~\bibnamefont
  {Marzari}}, \bibinfo {author} {\bibfnamefont {A.~A.}\ \bibnamefont
  {Mostofi}}, \bibinfo {author} {\bibfnamefont {J.~R.}\ \bibnamefont {Yates}},
  \bibinfo {author} {\bibfnamefont {I.}~\bibnamefont {Souza}},\ and\ \bibinfo
  {author} {\bibfnamefont {D.}~\bibnamefont {Vanderbilt}},\ }\href
  {https://doi.org/10.1103/RevModPhys.84.1419} {\bibfield  {journal} {\bibinfo
  {journal} {Rev. Mod. Phys.}\ }\textbf {\bibinfo {volume} {84}},\ \bibinfo
  {pages} {1419} (\bibinfo {year} {2012})}\BibitemShut {NoStop}%
\bibitem [{\citenamefont {Mostofi}\ \emph {et~al.}(2014)\citenamefont
  {Mostofi}, \citenamefont {Yates}, \citenamefont {Pizzi}, \citenamefont {Lee},
  \citenamefont {Souza}, \citenamefont {Vanderbilt},\ and\ \citenamefont
  {Marzari}}]{Mostofi-2014}%
  \BibitemOpen
  \bibfield  {author} {\bibinfo {author} {\bibfnamefont {A.~A.}\ \bibnamefont
  {Mostofi}}, \bibinfo {author} {\bibfnamefont {J.~R.}\ \bibnamefont {Yates}},
  \bibinfo {author} {\bibfnamefont {G.}~\bibnamefont {Pizzi}}, \bibinfo
  {author} {\bibfnamefont {Y.-S.}\ \bibnamefont {Lee}}, \bibinfo {author}
  {\bibfnamefont {I.}~\bibnamefont {Souza}}, \bibinfo {author} {\bibfnamefont
  {D.}~\bibnamefont {Vanderbilt}},\ and\ \bibinfo {author} {\bibfnamefont
  {N.}~\bibnamefont {Marzari}},\ }\href
  {https://doi.org/https://doi.org/10.1016/j.cpc.2014.05.003} {\bibfield
  {journal} {\bibinfo  {journal} {Computer Physics Communications}\ }\textbf
  {\bibinfo {volume} {185}},\ \bibinfo {pages} {2309 } (\bibinfo {year}
  {2014})}\BibitemShut {NoStop}%
\bibitem [{\citenamefont {Hanke}\ \emph {et~al.}(2015)\citenamefont {Hanke},
  \citenamefont {Freimuth}, \citenamefont {Bl\"ugel},\ and\ \citenamefont
  {Mokrousov}}]{Hanke-2015}%
  \BibitemOpen
  \bibfield  {author} {\bibinfo {author} {\bibfnamefont {J.-P.}\ \bibnamefont
  {Hanke}}, \bibinfo {author} {\bibfnamefont {F.}~\bibnamefont {Freimuth}},
  \bibinfo {author} {\bibfnamefont {S.}~\bibnamefont {Bl\"ugel}},\ and\
  \bibinfo {author} {\bibfnamefont {Y.}~\bibnamefont {Mokrousov}},\ }\href
  {https://doi.org/10.1103/PhysRevB.91.184413} {\bibfield  {journal} {\bibinfo
  {journal} {Phys. Rev. B}\ }\textbf {\bibinfo {volume} {91}},\ \bibinfo
  {pages} {184413} (\bibinfo {year} {2015})}\BibitemShut {NoStop}%
\bibitem [{\citenamefont {Freimuth}\ \emph {et~al.}(2008)\citenamefont
  {Freimuth}, \citenamefont {Mokrousov}, \citenamefont {Wortmann},
  \citenamefont {Heinze},\ and\ \citenamefont {Bl\"ugel}}]{Freimuth-2008}%
  \BibitemOpen
  \bibfield  {author} {\bibinfo {author} {\bibfnamefont {F.}~\bibnamefont
  {Freimuth}}, \bibinfo {author} {\bibfnamefont {Y.}~\bibnamefont {Mokrousov}},
  \bibinfo {author} {\bibfnamefont {D.}~\bibnamefont {Wortmann}}, \bibinfo
  {author} {\bibfnamefont {S.}~\bibnamefont {Heinze}},\ and\ \bibinfo {author}
  {\bibfnamefont {S.}~\bibnamefont {Bl\"ugel}},\ }\href
  {https://doi.org/10.1103/PhysRevB.78.035120} {\bibfield  {journal} {\bibinfo
  {journal} {Phys. Rev. B}\ }\textbf {\bibinfo {volume} {78}},\ \bibinfo
  {pages} {035120} (\bibinfo {year} {2008})}\BibitemShut {NoStop}%
\bibitem [{\citenamefont {Ziese}\ \emph {et~al.}(2010)\citenamefont {Ziese},
  \citenamefont {Vrejoiu},\ and\ \citenamefont {Hesse}}]{MAE-Ziese-2010}%
  \BibitemOpen
  \bibfield  {author} {\bibinfo {author} {\bibfnamefont {M.}~\bibnamefont
  {Ziese}}, \bibinfo {author} {\bibfnamefont {I.}~\bibnamefont {Vrejoiu}},\
  and\ \bibinfo {author} {\bibfnamefont {D.}~\bibnamefont {Hesse}},\ }\href
  {https://doi.org/10.1103/PhysRevB.81.184418} {\bibfield  {journal} {\bibinfo
  {journal} {Phys. Rev. B}\ }\textbf {\bibinfo {volume} {81}},\ \bibinfo
  {pages} {184418} (\bibinfo {year} {2010})}\BibitemShut {NoStop}%
\bibitem [{\citenamefont {Wang}\ \emph {et~al.}(2006)\citenamefont {Wang},
  \citenamefont {Yates}, \citenamefont {Souza},\ and\ \citenamefont
  {Vanderbilt}}]{Wang-Souza}%
  \BibitemOpen
  \bibfield  {author} {\bibinfo {author} {\bibfnamefont {X.}~\bibnamefont
  {Wang}}, \bibinfo {author} {\bibfnamefont {J.~R.}\ \bibnamefont {Yates}},
  \bibinfo {author} {\bibfnamefont {I.}~\bibnamefont {Souza}},\ and\ \bibinfo
  {author} {\bibfnamefont {D.}~\bibnamefont {Vanderbilt}},\ }\href
  {https://link.aps.org/doi/10.1103/PhysRevB.74.195118} {\bibfield  {journal}
  {\bibinfo  {journal} {Phys. Rev. B}\ }\textbf {\bibinfo {volume} {74}},\
  \bibinfo {pages} {195118} (\bibinfo {year} {2006})}\BibitemShut {NoStop}%
\bibitem [{\citenamefont {Yao}\ \emph {et~al.}(2004)\citenamefont {Yao},
  \citenamefont {Kleinman}, \citenamefont {MacDonald}, \citenamefont {Sinova},
  \citenamefont {Jungwirth}, \citenamefont {Wang}, \citenamefont {Wang},\ and\
  \citenamefont {Niu}}]{Yao-2004}%
  \BibitemOpen
  \bibfield  {author} {\bibinfo {author} {\bibfnamefont {Y.}~\bibnamefont
  {Yao}}, \bibinfo {author} {\bibfnamefont {L.}~\bibnamefont {Kleinman}},
  \bibinfo {author} {\bibfnamefont {A.~H.}\ \bibnamefont {MacDonald}}, \bibinfo
  {author} {\bibfnamefont {J.}~\bibnamefont {Sinova}}, \bibinfo {author}
  {\bibfnamefont {T.}~\bibnamefont {Jungwirth}}, \bibinfo {author}
  {\bibfnamefont {D.-s.}\ \bibnamefont {Wang}}, \bibinfo {author}
  {\bibfnamefont {E.}~\bibnamefont {Wang}},\ and\ \bibinfo {author}
  {\bibfnamefont {Q.}~\bibnamefont {Niu}},\ }\href
  {https://doi.org/10.1103/PhysRevLett.92.037204} {\bibfield  {journal}
  {\bibinfo  {journal} {Phys. Rev. Lett.}\ }\textbf {\bibinfo {volume} {92}},\
  \bibinfo {pages} {037204} (\bibinfo {year} {2004})}\BibitemShut {NoStop}%
\bibitem [{\citenamefont {Samanta}\ \emph {et~al.}(2020)\citenamefont
  {Samanta}, \citenamefont {Ležaić}, \citenamefont {Merte}, \citenamefont
  {Freimuth}, \citenamefont {Blügel},\ and\ \citenamefont
  {Mokrousov}}]{JAP-2020}%
  \BibitemOpen
  \bibfield  {author} {\bibinfo {author} {\bibfnamefont {K.}~\bibnamefont
  {Samanta}}, \bibinfo {author} {\bibfnamefont {M.}~\bibnamefont {Ležaić}},
  \bibinfo {author} {\bibfnamefont {M.}~\bibnamefont {Merte}}, \bibinfo
  {author} {\bibfnamefont {F.}~\bibnamefont {Freimuth}}, \bibinfo {author}
  {\bibfnamefont {S.}~\bibnamefont {Blügel}},\ and\ \bibinfo {author}
  {\bibfnamefont {Y.}~\bibnamefont {Mokrousov}},\ }\href
  {https://doi.org/10.1063/5.0005017} {\bibfield  {journal} {\bibinfo
  {journal} {Journal of Applied Physics}\ }\textbf {\bibinfo {volume} {127}},\
  \bibinfo {pages} {213904} (\bibinfo {year} {2020})},\ \Eprint
  {https://arxiv.org/abs/https://doi.org/10.1063/5.0005017}
  {https://doi.org/10.1063/5.0005017} \BibitemShut {NoStop}%
\end{thebibliography}%




\end{document}